\documentclass[%
 reprint,
superscriptaddress,
nofootinbib,
 amsmath,amssymb,
 aps,
]{revtex4-2}
\usepackage{multirow}
\usepackage{subfigure}
\usepackage{graphicx}
\usepackage{dcolumn}
\usepackage{bm}
\usepackage[colorlinks,
            linkcolor=red,
            anchorcolor=blue,
            citecolor=blue
            ]{hyperref}

\graphicspath{ {figure/} }
\begin{document}

\title{Interpretation of AMS-02 beryllium isotope fluxes using data-driven production cross sections}

\author{Meng-Jie Zhao}
\email{zhaomj@ihep.ac.cn}
 \affiliation{%
 Key Laboratory of Particle Astrophysics, Institute of High Energy Physics, Chinese Academy of Sciences, Beijing 100049, China}
\affiliation{
China Center of Advanced Science and Technology, Beijing 100190, China 
}%
 \author{Xiao-Jun Bi}
 \email{bixj@ihep.ac.cn}
\affiliation{%
 Key Laboratory of Particle Astrophysics, Institute of High Energy Physics, Chinese Academy of Sciences, Beijing 100049, China}
\affiliation{
 University of Chinese Academy of Sciences, Beijing 100049, China 
}%
\author{Kun Fang}
\email{fangkun@ihep.ac.cn}
\affiliation{%
 Key Laboratory of Particle Astrophysics, Institute of High Energy Physics, Chinese Academy of Sciences, Beijing 100049, China}
 \author{Peng-Fei Yin}
\email{yinpf@ihep.ac.cn}
\affiliation{%
 Key Laboratory of Particle Astrophysics, Institute of High Energy Physics, Chinese Academy of Sciences, Beijing 100049, China}



\date{\today}

\begin{abstract}
The Be isotopic measurements preliminarily reported by the AMS-02 Collaboration have reached an unprecedented energy of 12 GeV/$n$. As secondary cosmic rays (CRs), the Be isotopes include both stable and unstable species, which are crucial for constraining the propagation parameters of Galactic CRs. However, uncertainties in their production cross sections can skew the interpretation of the CR data, especially when cross-section measurements are of significantly lower quality than CR measurements. In this work, we consider the uncertainties of the cross sections to interpret the Be isotopic data by adopting a cross-section parametrization that fully utilizes the available experimental data. Owing to the high-quality measurements of the $^7$Be production cross section, we innovatively employ $^7$Be instead of $^9$Be to constrain propagation parameters. Notably, the diffusion halo thickness is constrained to $5.67\pm0.76$~kpc, representing a moderate value compared to previous analogous works. Combining the well-constrained CR propagation model and the precise CR measurements of $^9$Be, we conversely constrain the major production cross section of $^9$Be and find that it ought to be remarkably lower than previously thought. Our analysis also questions the reliability of certain cross sections measured by some experiments, potentially marking the first time CR data has been used to identify dubious nucleon production cross sections. The method presented in this work holds promise for analyzing upcoming isotopic data from other nuclei.
\end{abstract}
\maketitle


\section{\label{sec:level1}INTRODUCTION}

The study of Galactic cosmic rays (CRs) has advanced into an era of precision, owing to the improvements in experiment instruments and analysis methods. Exemplary experiments such as PAMELA \cite{Adriani:2011cu} and AMS-02 \cite{Aguilar:2018njt,Consolandi:2016fhd,Aguilar:2017hno,Aguilar:2021tos,AMS:2021brg} have achieved remarkable precision levels, with uncertainties now reduced to an impressive 1–3\%. These collaborations are collecting large numbers of CR events to explore more spectrum features that have not been discovered yet, which could update our knowledge on the CR acceleration and propagation mechanism, the spatial distribution of the CR sources and the interstellar medium (ISM), and the inelastic and production cross sections of nuclei reactions. The calculated background information can conversely provide reliable identification of signatures of the dark matter or new physics.

The measurement of secondary CR species is pivotal in constraining the propagation characteristics of Galactic CRs \cite{Strong:2007nh}. Secondary CRs arise from the fragmentation of heavier nuclei upon collision with the ISM gas. Consequently, the flux ratio of secondaries to their primary nuclei (e.g., B/C) informs us about the grammages of CRs, which represent the integrated gas density along their path before they escape from the Galaxy. Additionally, the ratio of unstable-to-stable secondary isotopes (e.g., $^{10}$Be/$^9$Be) provides insight into the residence time of CRs within the Galaxy. By synthesizing these two ratio types, we can independently determine the average diffusion coefficient for Galactic CRs and the height of the diffusion halo, which are the essential propagation parameters \cite{Strong:1998pw, Maurin:2001sj}.

Another crucial factor affecting the flux of secondary CRs is their production cross section. Regrettably, the precision and energy range of current cross-section measurements falls short of matching the secondary CR data obtained from space experiments. Most of the cross-section data were measured a long time ago, and the uncertainties for important reactions range from 10\% to 20\% \cite{Genolini:2018ekk,2023arXiv230706798G}, much larger than those found in high-precision CR datasets. Besides, for some astrophysically significant reaction channels, cross-section measurements are limited to low energies or are entirely lacking. In these cases, semi-empirical parametrizations have to be employed to approximate the cross sections \cite{Webber:1990pr,Silberberg:1998lxa,Summerer:1999ae}. The determination of cross sections is further complicated by the intricate network of multi-step and ghost-nuclei reactions. Nonetheless, the analysis can be simplified by ranking the cross-section production contribution, focusing primarily on the predominant channels \cite{Genolini:2018ekk,2023arXiv230706798G}. To sum up, the uncertainty in production cross sections is an essential consideration when using secondary species measurements to investigate CR propagation.

Li, Be, and B have the highest abundance among secondary CRs and have the most precise measurements \cite{Aguilar:2021tos}. Working under the assumption that these species are purely secondary in origin, the AMS-02 measurements can be understood within the conventional framework of CR propagation \cite{Weinrich:2020cmw, Boschini:2020jty, Yuan:2018lmc, DeLaTorreLuque:2021yfq,delaTorreLuque:2022vhm, Korsmeier:2021brc, Luque:2021nxb,Maurin:2022irz}. However, to account for certain anomalies, such as the overprediction of Be at low energies, minor adjustments to the production cross-sections within their uncertainties have been proposed \cite{Boschini:2019gow,delaTorreLuque:2022vhm}. On the other hand, by making use of the preliminary measurement of Be isotopes reported by the AMS-02 Collaboration \cite{AMS_icrc2021, AMS_ichep}, Refs.~\cite{2022arXiv220413085L,Jacobs:2023zch} argued that non-trivial CR propagation models might be necessary to account for the latest $^{10}$Be/$^9$Be, such as a two-zone diffusion model on a Galactic scale \cite{Tomassetti:2012ga,PhysRevD.104.123001}. It is evident that when experimental measurements of secondary CRs diverge from model predictions, disentangling whether this inconsistency arises from propagation factors or cross-section factors becomes a challenging problem.

In this study, we conduct the first individual analysis of the Be isotopes ($^7$Be, $^9$Be, and $^{10}$Be) detected by AMS-02 \cite{AMS_icrc2021, AMS_ichep}, moving beyond the traditional approach of solely interpreting the $^{10}$Be/$^9$Be ratio. $^7$Be (in the interstellar environment) and $^9$Be can be considered stable isotopes. Their propagation processes can be described by the same framework, which does not exhibit significant variability in fragmentation timescales or diffusion coefficients due to differences in atomic mass numbers \cite{Johannesson:2016rlh,Zhao:2022bon}. Therefore, should there be any discrepancies in the effectiveness of a propagation model in explaining the behavior of $^7$Be and $^9$Be, these must arise from uncertainties in the production cross sections, thus providing a good opportunity to disentangle propagation factors from production cross-section factors. Intriguingly, the cross-section measurements for $^7$Be are of significantly superior quality compared to those for $^9$Be. Therefore, we can initially use the CR energy spectrum of $^7$Be to constrain the propagation parameters. Then, using the derived propagation parameters, we can inversely constrain the production cross section of $^9$Be by interpreting its precise CR energy spectrum. Moreover, since the cross-section measurements for $^7$Be are of higher quality, we can break the degeneracy between the diffusion coefficient and the thickness of the diffusion halo by interpreting the CR measurements of $^7$Be and $^{10}$Be, superseding the previous method that used the $^{10}$Be/$^9$Be ratio. In the process of constraining the propagation parameters, we take into account the uncertainties of the production cross sections using existing cross-section data, also adding some updated data that offer enhanced precision and extend to higher energies (e.g., \cite{Amin:2021oow, Amin:2023fki}). 

This paper is organized as follows. In Sec.~\ref{sec:level2}, we introduce the CR propagation model and the cross-section setup for analysis. As a preliminary step, we estimate the Gaussian-distributed normalization for each reaction channel by fitting the cross-section data. In Sec.~\ref{sec:result}, we perform a combined fit to the CR data and the data-driven normalizations of the cross sections obtained earlier and discuss the results of the combined fit. CR data of $^7$Be, $^{10}$Be, B, C, and O are included in the fit. Same as $^7$Be, B is also an important stable secondary species, and we will show that its data can be explained within the same propagation model. The inclusion of B data thus enhances the constraints on propagation parameters. C and O, the dominant primary particles for these secondary CRs, must also be included in the fitting process. In Sec.~\ref{sec:be/b}, we compare the predicted $^9$Be and Be/B with the measured data to constrain the cross section of specific channels. Since the measurement of the Be/B ratio has been extended to 1 TeV, which is well above the energy range of measurements for Be isotopes, the inclusion of this data allows us to better distinguish between different high-energy extrapolation forms of the production cross sections. Finally, Sec.~\ref{sec:conclusion} is the summary of our findings above. For improved readability, we have moved several figures and discussions regarding the updated cross-section data and parametrizations used in this study to Appendix \ref{app:xsdata}.

\section{\label{sec:level2}CALCULATION SETUP}
\subsection{Propagation}
We adopt the standard CR propagation model with reacceleration, which is frequently used in CR analysis \cite{Trotta:2010mx, Genolini:2019ewc, Yuan:2018vgk}. Generally, the propagation equation of Galactic CRs is expressed as
\begin{eqnarray}
	 {\frac{\partial \psi}{\partial t}}=&&q(x,p)+\nabla\cdot(D_{xx}\nabla\psi-V_c\psi)
	 +{\frac{\partial}{\partial p}}[p^2D_{pp}{\frac{\partial}{\partial p}}({\frac{\psi}{p^2}})]\nonumber \\\label{eq:trans1}
	 &&-{\frac{\partial}{\partial p}}[\dot p\psi-{\frac{p}{3}}(\nabla\cdot V_c)\psi]
	-{\frac{\psi}{\tau_f}}-{\frac{\psi}{\tau_r}}\,, \label{eq:trans2}
\end{eqnarray}
where $\psi$ is the density per unit of particle momentum, $q(x,p)$ is the source distribution, $D_{xx}$ is the spatial diffusion coefficient, $V_c$ is the convection velocity\footnote{The effect of convection is ignored in the work for simplicity, as was found not necessary \cite{Yuan:2018vgk, Weinrich:2020cmw} for CR propagation.}, $D_{pp}$ is the momentum space diffusion coefficient, $\dot p\equiv dp/dt$ is the ionization and Coulomb losses terms, $\tau_f$ is the time scales for particle fragmentation, and $\tau_r$ is the time scales for radioactive decay.

The scattering of CR particles on randomly moving magnetohydrodynamics waves leads to stochastic acceleration, which is described in the transport equation as diffusion in momentum space $D_{pp}$. Considering the scenario where the CRs are reaccelerated by colliding with the interstellar random weak hydrodynamic waves, the relation between the spatial diffusion coefficient $D_{xx}$ and the momentum diffusion coefficient $D_{pp}$ is expressed as \cite{1994ApJ...431..705S}:

\begin{equation}
	D_{xx}D_{pp}=\frac{4p^2V_a^2}{3\delta(4-\delta)(4-\delta^2)\omega}\,.
\end{equation}

To solve the propagation equation, we adopt the numerical {\footnotesize GALPROP} v56\footnote{The current version is available at \url{https://galprop.stanford.edu}.} framework  \cite{Strong:1998pw,Strong:1998fr}.
The code of {\footnotesize GALPROP} includes a nuclear reaction network to calculate the 1-step and multi-step fragmentation of the particles that collide on the interstellar medium gas.
The reaction network is a series of repeated procedures, starting at solving the propagation equation of the heaviest nuclei $\rm{^{64}_{28}Ni}$, computing all the resulting secondary sources, and then proceeding to the nuclei with $A-1$, where $A$ is the mass number of nuclei. 
This network makes use of the available cross-section measurements and also parametrizations, which are significant for predicting the cross sections of the data-lacking channels.
In this work, we have made several updates to the nuclear data and adjusted the parametrizations of some dominant channels based on recent observations. See Appendix~\ref{app:xsdata} for more details.

The high-energy spectral hardening has been discovered by several observations \cite{Adriani:2011cu,Consolandi:2016fhd,Aguilar:2017hno,Aguilar:2018njt,Panov:2011ak,Adriani:2019aft,Adriani:2020wyg,An:2019wcw,Alemanno_2021,Yoon:2017qjx}, but its origin is still under debate \cite{Ptuskin:2012qu,Blasi:2012yr,Thoudam:2011aa,Ohira:2010eq,PhysRevD.104.123001}. 
For simplicity, the diffusion coefficient $D_{xx}$ is defined as a broken power-law with a harder index $\delta_h$ above the high-energy break rigidity $R_h$.
We performed a prior fitting for the B/C and B/O ratios measured by AMS-02 and DAMPE \cite{Aguilar:2018njt, Collaboration:2022vwu} to pin down the break rigidity and the slope index change, which are $R_h=280$ GV and $\delta_h=\delta-0.226$. This can reduce the parameters in the following fitting process.

To describe the common bump structure of nuclei fluxes at low rigidities \cite{Johannesson:2016rlh, Phan:2021iht}, we model the injection spectrum of primary nuclei as a broken power law, with slope indices $\nu_0$ and $\nu_1$ below and above the low-energy break rigidity $R_{br}$: 
\begin{equation}
\begin{aligned}
   q(R)=\begin{cases} (R/R_{br})^{\nu_0}\,,  \quad& R<R_{br} \\
   (R/R_{br})^{\nu_1}\,,  \quad& R\geq R_{br}\end{cases}\,.
\end{aligned}
\end{equation}
We assume that the carbon, nitrogen, and oxygen nuclei share the same injection parameters ($R_{br}$, $\nu_0$ and $\nu_1$) as suggested by the observation of AMS-02 \cite{Aguilar:2017hno, Aguilar:2021tos}. The source abundance $A_C$ and $A_O$ are also free parameters. Other individual isotopic source abundances and slope indices have been adjusted and fixed to match the corresponding primary fluxes \cite{Consolandi:2016fhd, Aguilar:2017hno, AMS:2020cai, Aguilar:2021tos, AMS:2021brg, AMS:2021lxc}.

The B/C ratio is constrained by the AMS-02 measurements, and the degeneracy of $L/D_{xx}$ is kept as well. To break the degeneracy, the $\rm^{10}Be/^{9}Be$ ratio is often adopted. The radioactive decay of unstable $\rm^{10}Be$ is related to the decay timescale and is sensitive to the Galactic halo height. If the halo size increases, the $\rm^{10}Be/^{9}Be$ and $\rm^{10}Be/^{7}Be$ ratio become smaller.
Note that the measurements of $\rm^{9}Be$ flux and $\rm^{10}Be/^{9}Be$ ratio are not used in the fitting. As shown in Appendix~\ref{app:xsdata}, the uncertainty of $\rm^{9}Be$ production is much larger than that of $\rm^{7}Be$ and $\rm^{10}Be$, owing to less observation available. Prior fitting implies that the default cross-section parametrization predicts a systematically higher $\rm^{9}Be$ flux relative to the CR measurements of AMS-02. This anomaly will be further analyzed and discussed in Sec.~\ref{sec:be/b}.
The cross-section uncertainty of $\rm^{7}Be$ is smaller, providing better constraining quality than that of $\rm^{9}Be$.
Therefore, in the work, we constrain the Galactic halo height by $\rm^{7}Be$ and $\rm^{10}Be$ instead.

In Table~\ref{tab:exp}, we list all the CR measurements used for constraining the parameters in the work. The precise measurements of C, O, and B fluxes from AMS-02 \cite{AMS:2023anq} are used, and other data are included for better parameter constraints. For the C and O fluxes, we also include the CALET \cite{Adriani:2020wyg}, NUCLEON  \cite{Gorbunov:2018stf}, and CREAM-II \cite{Ahn:2009tb} measurements to cover the multi-TeV energy region. The isotope $\rm^{7}Be$ and $\rm^{10}Be$ fluxes are constrained by the recently reported preliminary AMS-02 data \cite{AMS_ichep}.

\begin{table*}
\caption{\label{tab:exp}Data Used in This Analysis.}
\begin{ruledtabular}
\begin{tabular}{cccc}
 Experiment&Energy Range&data points&Reference\\ \hline
 \multicolumn{4}{c}{\textbf{C}}\\
  NUCLEON(7/2015-6/2017)& 250-17000 GeV/$n$& 10 & \cite{Gorbunov:2018stf}\\
 CREAM-II(12/2005-1/2006)&  85-7500 GeV/$n$& 9 & \cite{Ahn:2009tb}\\
  CALET(10/2015-10/2019)×1.27\footnote{A multiplication of 1.27 is needed for the C and O measurements of CALET to achieve alignment with AMS-02 \cite{Adriani:2020wyg}.}& 10-1700 GeV/$n$& 22 & \cite{Adriani:2020wyg}\\
 AMS-02(5/2011-5/2021)& 2-2000 GV& 66 & \cite{AMS:2023anq}\\
 Voyager 1-HET(2012-2015)&  0.02-0.13 GeV/$n$& 8 & \cite{Cummings:2016pdr}\\
 \multicolumn{4}{c}{\textbf{O}}\\
  NUCLEON(7/2015-6/2017)& 300-13000 GeV/$n$& 9 & \cite{Gorbunov:2018stf}\\
 CREAM-II(12/2005-1/2006)&  64-7500 GeV/$n$& 9 & \cite{Ahn:2009tb}\\
  CALET(10/2015-10/2019)×1.27 & 10-1700 GeV/$n$& 22 & \cite{Adriani:2020wyg}\\
 AMS-02(5/2011-5/2021)& 2-2000 GV& 66 & \cite{AMS:2023anq}\\
 Voyager 1-HET(2012-2015)&  0.02-0.15 GeV/$n$& 10 & \cite{Cummings:2016pdr}\\
  \multicolumn{4}{c}{\textbf{B}}\\
 AMS-02(5/2011-5/2021)& 2-2000 GV& 66 & \cite{AMS:2023anq}\\
Voyager 1-HET(2012-2015)&  0.02-0.11 GeV/$n$& 8 & \cite{Cummings:2016pdr}\\
 \multicolumn{4}{c}{$\mathbf{^{7}Be}$,$\mathbf{^{10}Be}$}\\
AMS-02(preliminary)&  0.7-11 GeV/$n$& 26 & \cite{AMS_ichep}\\
\end{tabular}
\end{ruledtabular}
\end{table*}
To consider the solar modulation effect on the spectrum inside the heliosphere, we adopt the force-field approximation \cite{Gleeson:1968zza}. The strength is described by the solar modulation potential $\phi$. The CR measurements used were taken during a similar period (B, C, and O taken during May 2011-May 2021  \cite{AMS:2023anq} and beryllium isotopes taken during May 2011-nearly 2022  \cite{AMS_ichep}), hence we assume that they share almost the same $\phi$. To estimate the modulation potential, the local interstellar spectra are also needed for determining the unmodulated ($\phi=0$) fluxes. Hence, we include the Voyager 1 measurements of B, C, and O fluxes \cite{Cummings:2016pdr} in the fitting.

\subsection{Cross Section}
The cross-section uncertainty has been regarded as a subdominant or even negligible factor in CR research for a long time due to the larger uncertainties of CR fluxes.
However, as CR measurements become increasingly precise, the uncertainty in the production cross sections of secondary CRs has emerged as a critical factor that must be considered when constraining propagation parameters \cite{Weinrich:2020cmw, Weinrich:2020ftb}. Acknowledging its impact, we introduce additional parameters to adjust the cross sections of secondary nuclei during the fitting process.

To begin, it is essential to quantify the uncertainties associated with each production channel. Here we follow a data analysis routine introduced in our previous work \cite{Zhao:2022bon}, which was used to analyze the cross-section uncertainty of the F and B production. Through the implementation of a data-driven least-squares fitting approach, we can get the Gaussian-distributed dispersion values for the renormalization consistent with the selected dataset. 

The default parametrization developed in {\footnotesize GALPROP} code is [GAL12] ([GAL22]), partly based on their fits of a compilation of cross-section measurements and code evaluations, and partly based on the Webber's \cite{Webber:1990pr} (or Silberberg's \cite{Silberberg:1998lxa}) parametrization model with semi-empirical formulas.
For isotope Be and B production cross sections, the parametrization follows a direct fit to the available data. The plateau\footnote{Above which the cross section appears to be energy-independent.} is mostly decided by the highest-energy measurement in the data sheet named \texttt{eval\_iso\_cs.dat}, which has been updated in Appendix~\ref{app:xsdata} by adding more observations.
General features of nucleon production have been discovered by analyzing available measurements, for example, the cross section becomes constant at above $2~$GeV/$n$ \cite{Glagolev:1993ys}.
However, there is an energy-dependent rise observed in the total and inelastic nucleon-nucleon cross section \cite{Block:2011vz, Block:2015mjw}.
As implied from recent analysis \cite{Webber_2003, Reinert:2017aga}, a slow change of the cross section can still appear above $2~$GeV/$n$, making the determination of the plateau according to the highest-energy measurement less reliable.
We notice that these production channels lack of high-energy ($>10~$GeV/$n$) measurements, which means that the prediction of secondary nuclei relies seriously on the extrapolation of the mid-energy data measured by multiple experiments.
Therefore, the uncertainty in the extrapolation region should not be strictly constrained as well as at low energies.
Data points above $2~$GeV/$n$ should be important for analyzing and constraining the plateau, thus we choose to use high-energy data above that energy to determine the uncertainty in the extrapolation region.

\begin{table}
\caption{\label{tab:dispersion}The average and high-energy uncertainties of different cross-section channels for the production of B and Be isotopes.}
\begin{ruledtabular}
\begin{tabular}{ccc}
 Channel&$\omega_0$&$\omega_1$\\ \hline
 $^{12}\text{C} \longrightarrow ^{10}\text{B}$&0.005&0.113\\
 $^{12}\text{C} \longrightarrow ^{11}\text{B}$&0.009&0.066\\  
     $^{12}\text{C} \longrightarrow ^{10}\text{C} \longrightarrow ^{10}\text{B}$&0.012&0.081\\  
     $^{12}\text{C} \longrightarrow ^{11}\text{C} \longrightarrow ^{11}\text{B}$&0.003&0.011\\  
      $^{16}\text{O} \longrightarrow ^{10}\text{B}$&0.024&0.038\\
 $^{16}\text{O} \longrightarrow ^{11}\text{B}$&0.022&0.033\\  
     $^{16}\text{O} \longrightarrow ^{10}\text{C} \longrightarrow ^{10}\text{B}$&0.051&0.111\\  
     $^{16}\text{O} \longrightarrow ^{11}\text{C} \longrightarrow ^{11}\text{B}$&0.002&0.063\\  \hline
 $^{12}\text{C} \longrightarrow ^{7}\text{Be}$&0.004&0.020\\
 $^{12}\text{C} \longrightarrow ^{10}\text{Be}$&0.009&0.084\\  
 $^{16}\text{O} \longrightarrow ^{7}\text{Be}$&0.007&0.021\\
 $^{16}\text{O} \longrightarrow ^{10}\text{Be}$&0.013&0.079 
\end{tabular}
\end{ruledtabular}
\end{table}
We have made several least-squares fits to the cross-section data with the parametrization formulae given by the {\footnotesize GALPROP} code (see Appendix~\ref{app:xsdata}), to determine the Gaussian-distributed dispersion as an estimate of the uncertainty.
In Table~\ref{tab:dispersion} we list the fitting result of cross-section uncertainties. These channels are the main production channels for B, $\rm^{7}$Be and $\rm^{10}$Be, including the contributions from ghost nuclei\footnote{The short-lived intermediate nuclei that will decay quickly before they can collide on the ISM gas.} $\rm^{10}$C and $\rm^{11}$C.
The average uncertainty $\omega_0$ of the individual channel is determined by a global fitting for all of the available data, while the high-energy uncertainty $\omega_1$ is determined by exclusively fitting the data points at $\ge 2~$GeV/$n$.
We notice that the ghost nuclei $\rm^{10}$C contributes less than 5\% of the total B flux at all energies, and its uncertainties are 1 order of magnitude smaller, which should not significantly impact the resulting flux ($<0.5\%$). Thus, we ignore the uncertainty contributions from $\rm^{12}\text{C} \longrightarrow ^{10}\text{C} \longrightarrow ^{10}\text{B}$ and $\rm^{16}\text{O} \longrightarrow ^{10}\text{C} \longrightarrow ^{10}\text{B}$ channels.

An energy-dependent modification rather than a constant re-scaling is better for describing the cross-section uncertainties as listed in Table~\ref{tab:dispersion}.
Following Refs. \cite{Reinert:2017aga, Evoli:2019wwu}, we construct a two-part modification formula to renormalize the production cross section $\sigma$, which depends on the kinetic energy per nucleon $E_{kin/n}$:
\begin{eqnarray}\label{eq:sigma}
	\sigma=\sigma^{0}\left[1+\frac{\mu}{1+(E_{th}/E_{kin/n})^2}+\frac{\mu\omega_0/\omega_1}{1+(E_{kin/n}/E_{th})^2}\right]\,,
\end{eqnarray}
where $\sigma^0$ is the cross section determined by a data-driven parametrization introduced in Appendix~\ref{app:xsdata}, and $\mu$ is the renormalization factor that will act as a free parameter in the CR data fit.
The threshold energy $E_{th}$ is set to $2~$GeV/$n$, to allow for a slow change of the cross section above that energy \cite{Reinert:2017aga, Evoli:2019wwu}.

\begin{figure}[htbp]
\includegraphics[width=0.5\textwidth,trim=0 0 0 0,clip]{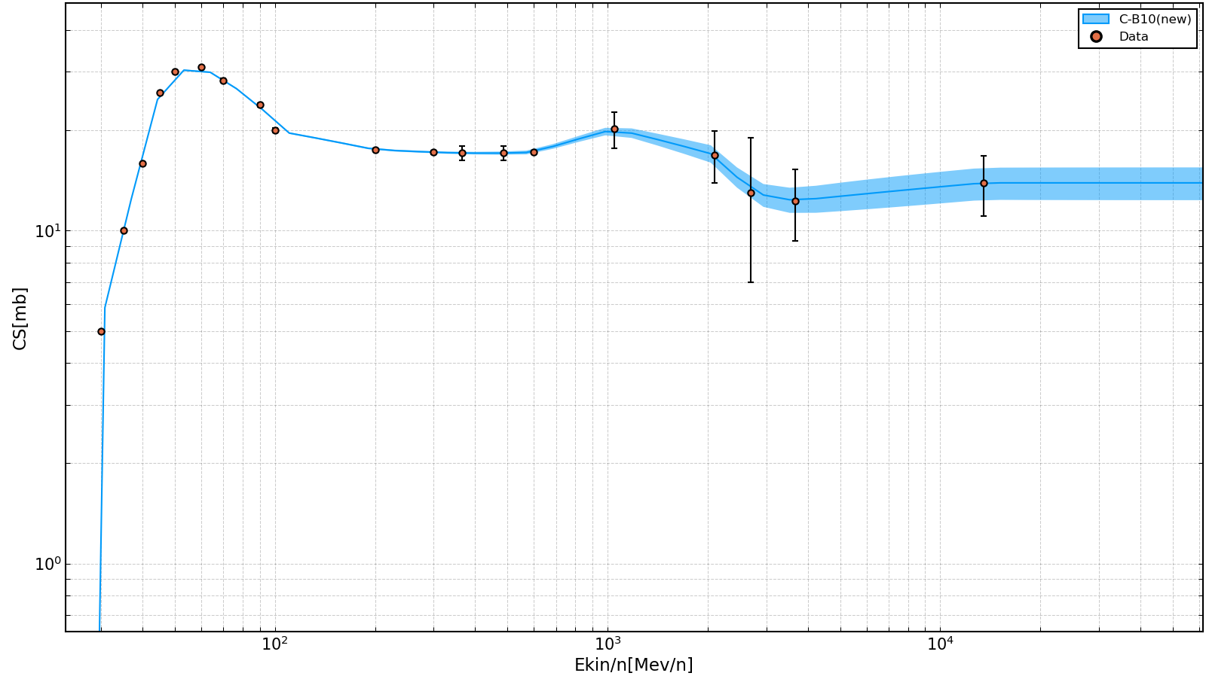}
\caption{\label{fig:two-part} 
Channels: $\rm C + p \longrightarrow ^{10}\text{B}$. The blue line is the cross section $\sigma_0$ obtained by the data-driven parametrization. The blue band is the $1\sigma$ confidence interval allowed by the cross-section measurements, the definition of which is referred to Eq.~(\ref{eq:sigma}).}
\end{figure}

In the high-energy extrapolation region, Eq.~(\ref{eq:sigma}) approaches the asymptotic value $\sigma\rightarrow\sigma^{0}(1+\mu)$. When the energy becomes lower, the asymptotic value tends to be smaller as $\sigma\rightarrow\sigma^{0}(1+\mu\omega_0/\omega_1)$. Thus, the factor $\mu$ represents the renormalization above $2~$GeV/$n$. When $\mu$ approaches $\omega_1$ ($-\omega_1$), the low-energy renormalization acts as narrower constraints, approaching $\omega_0$ ($-\omega_0$). As an example, we show the 68\% confidence interval of the parametrization renormalization for the $\rm C + p \longrightarrow ^{10}\text{B}$ channel in Fig.~\ref{fig:two-part}. Above $2~$GeV/$n$, the renormalization range approaches [$-\omega_1$, $\omega_1$].

The individual contributions from N, Ne, Mg, Si, and Fe to the concerned secondaries cannot be ignored, as the cumulative contributions are up to 20\% of the total fluxes. However, the cross-section measurements of these subdominant channels are relatively scarce, and the uncertainties are much larger than those of the dominant channels listed in Table~\ref{tab:dispersion}. To simply consider the uncertainties among subdominant channels, we renormalize the cross section of them by using Eq.~(\ref{eq:sigma}), while the renormalization factor $\mu$ is decided by the average renormalization of C and O as $(\mu_C+\mu_O)/2$ in the following fitting process.

\section{FITTING RESULTS\label{sec:result}}
For the goodness-of-fit of the model to the data, we use the least-$\chi^2$ method. The $\chi^2$ statistic is expressed as
\begin{equation}
\label{eq:chi}
  \chi^2=\sum\chi^2_{{\rm cr},q}+\chi^2_{\rm{cs}},
\end{equation}
\begin{equation}
\label{eq:data}
  \chi^2_{\rm cr}=\sum_{i=1}^{\rm{bin}}(\frac{y_i^{\rm{data}}-y_i^{\rm{model}}}{\sigma_i^{\rm{data}}})^2,
\end{equation}
\begin{equation}
\label{eq:chi_cs}
  \chi^2_{\rm{cs}}=\sum_{i=1}^{n_{\rm{cs}}}(\frac{\mu_i}{\omega_{1i}})^2,
\end{equation}
where $y_i^{\rm{model}}$ is the CR flux predicted by the model, and $y_i^{\rm{data}}$ and $\sigma_i^{\rm{data}}$ are the flux and error of the CR measurements, respectively.
In Eq.~\ref{eq:chi}, $q$ runs over the CR fluxes of B, C, O, $\rm^7$Be, and $\rm^{10}$Be (in Table~\ref{tab:exp}), each energy bin is calculated separately to get the quadratic distance between the data and the model.
The constraint from the cross-section data contributes an additional term $\chi^2_{\rm{cs}}$, where $\omega_{1i}$ is the high-energy uncertainty of the specific channel $i$ taken from Table~\ref{tab:dispersion}, and $\mu_i$ is the tested value of renormalization factor in the fit.
Note that the CR data contributes most of the degrees of freedom (d.o.f.) and has better precision than the cross-section data. This means that even if $\chi^2/n_{\rm{d.o.f.}}$ is less than 1, we still need to check whether $\chi^2_{\rm{cs}}/n_{\rm{cs}}$ is reasonable.

Markov chain Monte Carlo (MCMC) methods are widely used in Bayesian inference and are powerful to sample the multi-dimensional parameter space for CR propagation models  \cite{Masi:2016uby, Putze:2010zn, Yuan:2017ozr, Johannesson:2016rlh}.
We perform a combined fitting of CR observations together with cross-section uncertainties by using the public code {\footnotesize CosmoMC}\footnote{\url{https://cosmologist.info/cosmomc}.} as a generic Monte Carlo sampler to explore parameter space.
In the previous paper \cite{PhysRevD.104.123001}, we have introduced the basic settings for {\footnotesize GALPROP} and {\footnotesize CosmoMC}.
For the resolution of the {\footnotesize GALPROP} calculation in the work, we set a 2D spatial grid of $dr=1$~kpc and $dz=0.2$~kpc, and an energy grid of Ekin\_factor = 1.2, considering both accuracy and speed. The size of the initial time step (start\_timestep) is set to be 1.0e8, which is smaller than the default. We have checked that it does not affect the results. Other parameters are kept as the defaults of {\footnotesize GALPROP}~v56.

The group of free parameters in the fitting procedure is 
\begin{eqnarray*}
    \bm{\theta}=\{D_0,\delta,L,V_a,\eta,
    \nu_0,\nu_1,R_{br},A_C,A_O,\phi,\\
    \mu_{C-B_{10}},\mu_{C-B_{11}},\mu_{C-C_{11}},...\},
\end{eqnarray*}
where the $D_0$, $\delta$, $L$, $V_a$, and $\eta$ are the propagation parameters; the $\nu_0$,$\nu_1$,$R_{br}$,$A_C$,$A_O$ and $\phi$ are the injection parameters and the solar modulation potential, respectively; the rest parameters are the renormalization factors for the cross sections of specific channels.

The fitting result is shown in Table~\ref{tab:best}. The goodness of fit is acceptable with $\chi^2/n_{\rm{d.o.f.}}=232.6/320$, and the $\chi^2$ contribution from cross-section renormalization is $\chi^2_{\rm{cs}}/n_{\rm{cs}}=3.2/10$.
The best-fit slope index of diffusion in the halo $\delta=0.433$ prefers the Iroshnikov-Kraichnan type (1/2) \cite{Kraichnan:1965zz} hydromagnetic turbulence as the source of diffusion. 
The posterior distribution of the halo height $L$ is constrained to be $5.674\pm0.758~$kpc at 68\% confidence. The result is smaller, compared with the $\sim 7~$kpc result from some works \cite{Evoli:2019iih,delaTorreLuque:2022vhm}.
The difference mainly originates from the adopted cross-section parametrizations, since we have updated the cross section datasets and calculated a data-driven parametrization. We refer the reader to Appendix~\ref{app:xsdata} for checking the differences.
On the other hand, a smaller halo height is constrained as the result of some works \cite{Maurin:2022gfm}. The differences in parametrizations may also explain it, for example, the inclusion of [NA61/SHINE] data \cite{Amin:2021oow, Amin:2023fki} has increased the production of secondary B, which allows a slightly larger halo height for nuclei to diffuse out of the Galaxy.
Another reason is that the $\rm^{10}\text{Be}/^{9}\text{Be}$ ratio observed by AMS-02 \cite{AMS_ichep} is smaller than the data of ISOMAX \cite{Hams:2004rz}, indicating the current estimation of the $^{10}$Be flux is smaller than that obtained by earlier works based on the ISOMAX data.
As the diffusion should dominate the fluxes of $\rm^{10}\text{Be}$ over the radioactive decay at high energies, a smaller $^{10}$Be flux indicates a larger halo size.
The quality of the isotope fluxes measured by AMS-02 is better and can give a strict constraint, while the inclusion of cross-section uncertainty enlarges the confidence interval of the halo height conversely. As a result, the constraint of the halo height shown in Table~\ref{tab:best} is slightly larger in comparison with the $\pm0.4$~kpc result based on a simple analysis without considering cross-section uncertainties \cite{Jacobs:2023zch}.

\begin{table*}
\caption{\label{tab:best}The prior range, best-fit values, and posterior 95\% range of all parameters in the combined fitting.}
\begin{ruledtabular}
\begin{tabular}{cccc}
 Parameter&Prior range&Best-fit values&Posterior 95\% range\\ \hline
 $D_0(\rm10^{28}cm^2s^{-1})$&[0,15.0]&5.197&[4.176,6.396]\\
 $\delta$&[0.2,1.0]&0.433&[0.424,0.456]\\
 $L$(kpc)&[1.0,20.0]&5.674&[4.384,7.443]\\
 $V_a$(km/s)&[0,50]&15.409&[12.333,18.068]\\
 $\eta$&[-5,5]&-0.484&[-0.732,-0.161]\\
 $\nu_0$&[0.5,2.4]&1.249&[1.003,1.446]\\
 $\nu_1$&[2.2,2.5]&2.390&[2.372,2.400]\\
 $R_{\rm{br}}$(GV)&[0.1,15]&2.088&[1.743,2.551]\\
  $A_c(10^{3})$\footnote{The abundance of proton $A_p$ is 1.06×$10^6$, and the normalization of proton flux at 100 GeV is $4.4*10^{-9}\rm{cm}^{-2}\rm{s}^{-1}\rm{sr}^{-1}\rm{MeV}^{-1}$.}&[2.5,4.5]&3.304&[3.257,3.328]\\
 $A_o(10^{3})$&[3.5,5.5]&4.114&[4.062,4.185]\\
 $\phi$(GV)&[0.4,1.0]&0.645&[0.619,0.697]\\ \hline
 $\mu_{C-B10}$&[-0.5,0.5]&0.099&[-0.085,0.316]\\ 
 $\mu_{C-B11}$&[-0.5,0.5]&0.075&[-0.007,0.211]\\ 
 $\mu_{C-C11}$&[-0.5,0.5]&-0.001&[-0.018,0.025]\\ 
 $\mu_{O-B10}$&[-0.5,0.5]&0.010&[-0.062,0.081]\\ 
 $\mu_{O-B11}$&[-0.5,0.5]&0.026&[-0.058,0.071]\\ 
 $\mu_{O-C11}$&[-0.5,0.5]&0.026&[-0.095,0.134]\\ 
 $\mu_{C-Be7}$&[-0.5,0.5]&-0.001&[-0.037,0.049]\\ 
 $\mu_{C-Be10}$&[-0.5,0.5]&-0.039&[-0.159,0.122]\\ 
 $\mu_{O-Be7}$&[-0.5,0.5]&0.001&[-0.036,0.045]\\ 
 $\mu_{O-Be10}$&[-0.5,0.5]&0.031&[-0.154,0.128]\\ \hline
 $\chi^2_{\rm{min}}/n_{\rm{dof}}$&$\cdots$& 232.6/320 &$\cdots$\\
 $\chi^2_{\rm{cs}}/n_{\rm{cs}}$&$\cdots$& 3.2/10 &$\cdots$\\
\end{tabular}
\end{ruledtabular}
\end{table*}

\begin{figure}[htbp]
\includegraphics[width=0.48\textwidth,trim=0 0 0 0,clip]{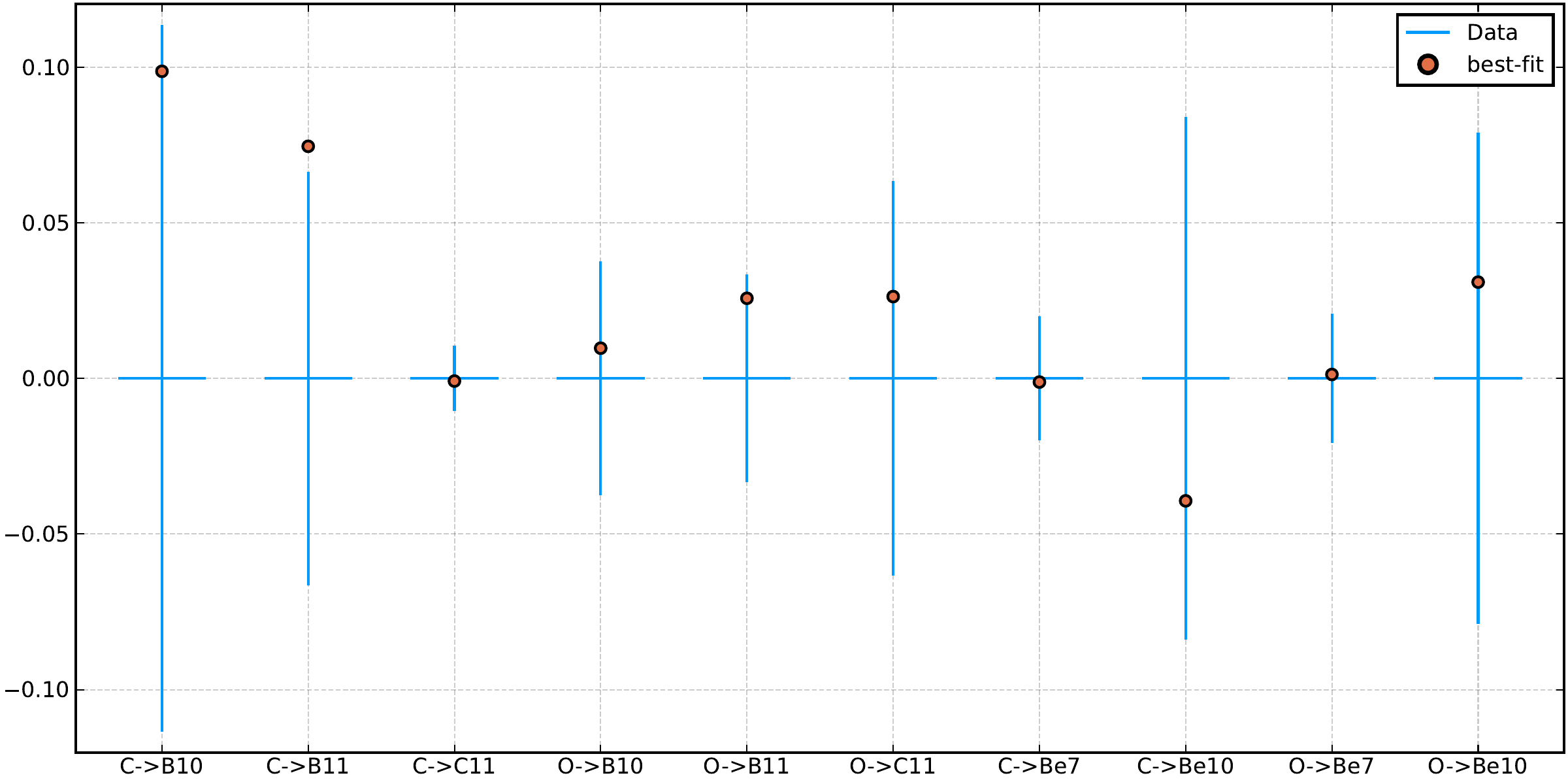}
\caption{\label{fig:xs-sys} 
Comparison between cross-section uncertainties ($\omega_1$ in Table~\ref{tab:dispersion}) and best-fit renormalization factors ($\mu$ in Table~\ref{tab:best}).}
\end{figure}

\begin{figure}[htbp]
\includegraphics[width=0.5\textwidth,trim=0 0 0 0,clip]{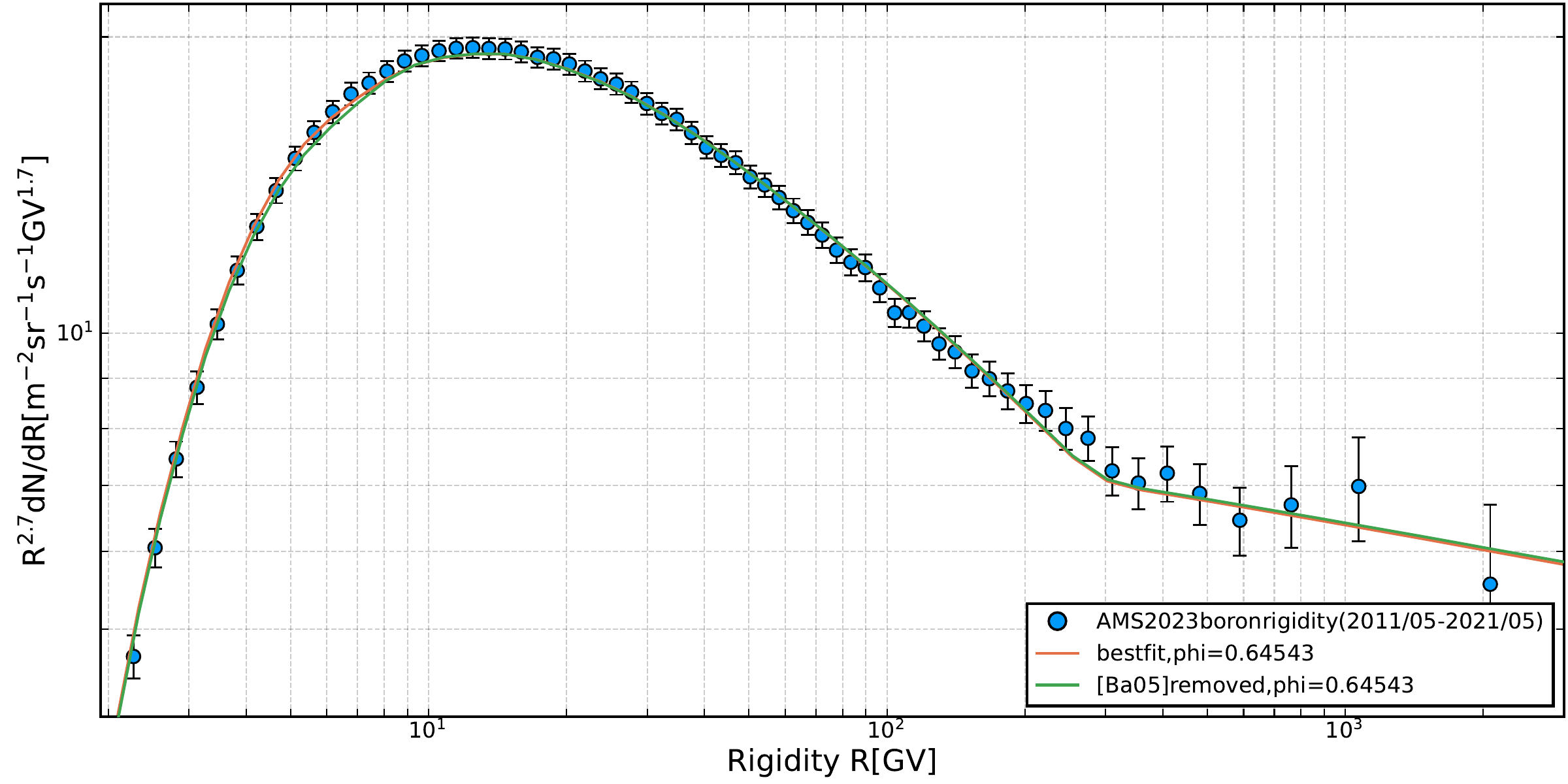}
\includegraphics[width=0.5\textwidth,trim=0 0 0 0,clip]{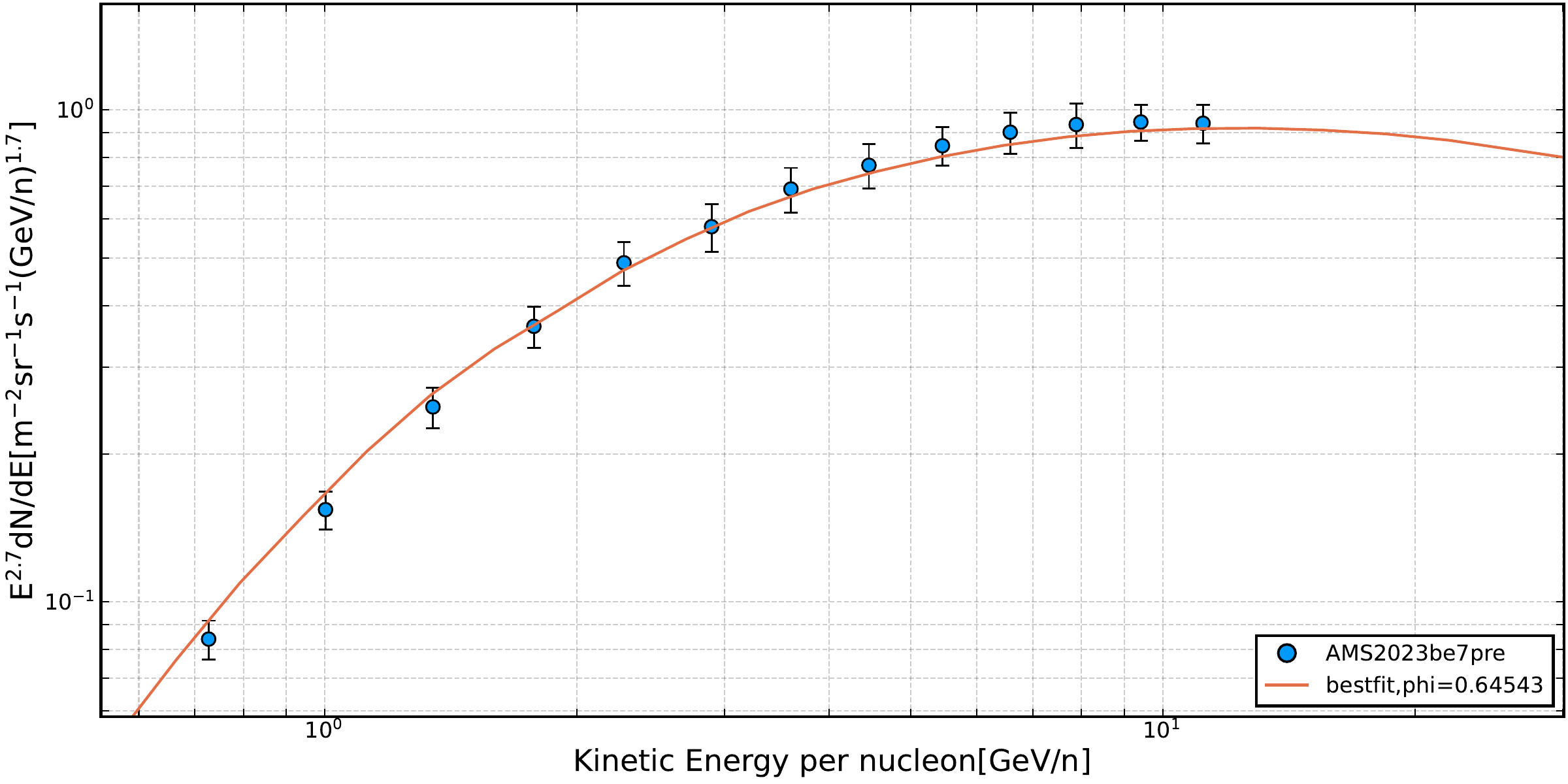}
\caption{\label{fig:boron-XS} 
Top panel: The B spectrum calculated with the best-fit parameters (green solid line) compared with the AMS-02 experimental data \cite{AMS:2023anq}. The red solid line represents the result without using the [Ba05] data and renormalizing the cross section.
Bottom panel: The $\rm^{7}\text{Be}$ spectrum calculated with the best-fit parameters compared with the AMS-02 experimental data \cite{AMS_ichep}.}
\end{figure}

In Fig.~\ref{fig:xs-sys}, we illustrate the best-fit parameters of the cross-section renormalization factor, compared with the data expectation ranges from Table~\ref{tab:dispersion}. 
The first 6 channels are related to the B production. The best-fit result implies a systematical overproduction to fulfill the CR observation, which may not be attributed to the random errors of the cross sections. 
As we mentioned in Sec.~\ref{sec:level2}, the data expectation $\omega_1$ is obtained by fitting the cross-section data points at $\geq2~$GeV/$n$. 
Coincidentally, we noticed that the cross-section measurements of [Ba05] \cite{Bazarov:2005qu} at 3.25 GeV/$n$ for both the two dominant channels of $\rm^{16}\text{O} \longrightarrow ^{11}\text{B}$ and $\rm^{16}\text{O} \longrightarrow ^{11}\text{C}$ imply a reduction of about 30\% compared with other measurements, which significantly changes the expectation of parametrizations at high energies.
By removing those two points and ignoring the renormalization of the first 6 channels ($\mu_{C-B_{10}},\mu_{C-B_{11}},\mu_{C-C_{11}},...$), we calculate another B flux illustrated in the top panel of Fig.~\ref{fig:boron-XS}. Compared with the best-fit result of the B spectrum, we find that they are almost the same, implying that the renormalization factors $\mu$ of B production channels are no longer required if the [Ba05] measurements are discarded. The renormalization of the cross section can also impact the fluxes below 2 GeV/$n$ and prefers an increment of B production at low energies. The removal of [Ba05] does not change the spectrum prediction below 2 GeV/$n$, since the cross section is constrained by other measurements. As a result, the slightly under-prediction of fluxes around 5 GV can be seen in the top panel, which needs a thorough estimation of the uncertainties from the cross section, solar modulation, and also the systematic errors of AMS-02 data with its covariance matrices \cite{Weinrich:2020cmw}.
We also illustrate the best-fit $\rm^{7}\text{Be}$ flux in the bottom panel of Fig.~\ref{fig:boron-XS}, which matches the AMS-02 experimental data well, without the need for significant correction to the production cross section of $^7$Be. As $\rm^{7}\text{Be}$ and B nuclei both originated from the fragmentation of primary CR nuclei (mostly C and O) and have similar mass numbers, they should indeed be explained by almost the same diffusion process in the Galaxy. Thus, our results indicate that the production cross-section measurements and CR flux measurements of $^7$Be and B are self-consistent.

\begin{figure}[htbp]
\includegraphics[width=0.5\textwidth,trim=0 0 0 0,clip]{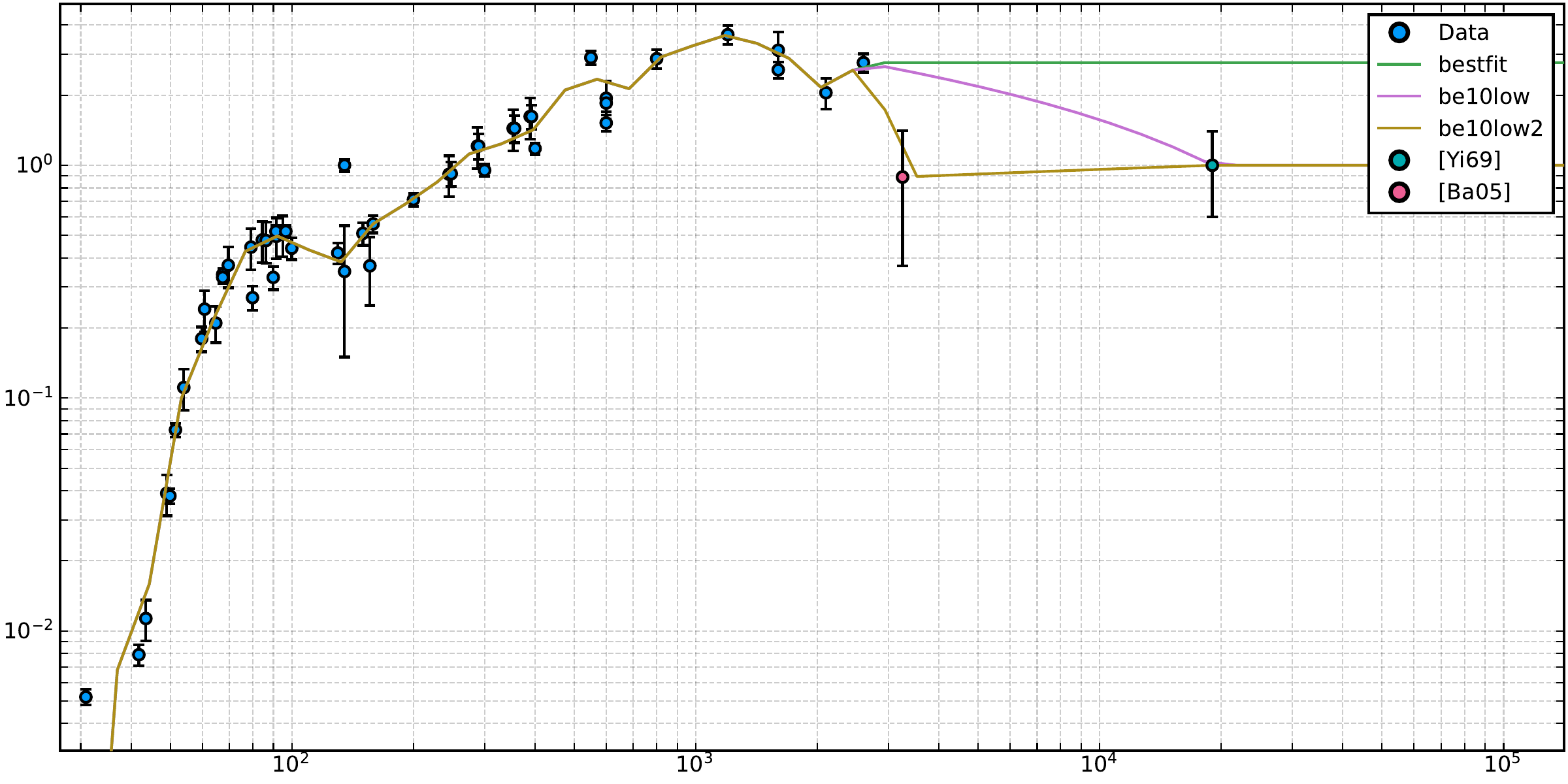}
\includegraphics[width=0.5\textwidth,trim=0 0 0 0,clip]{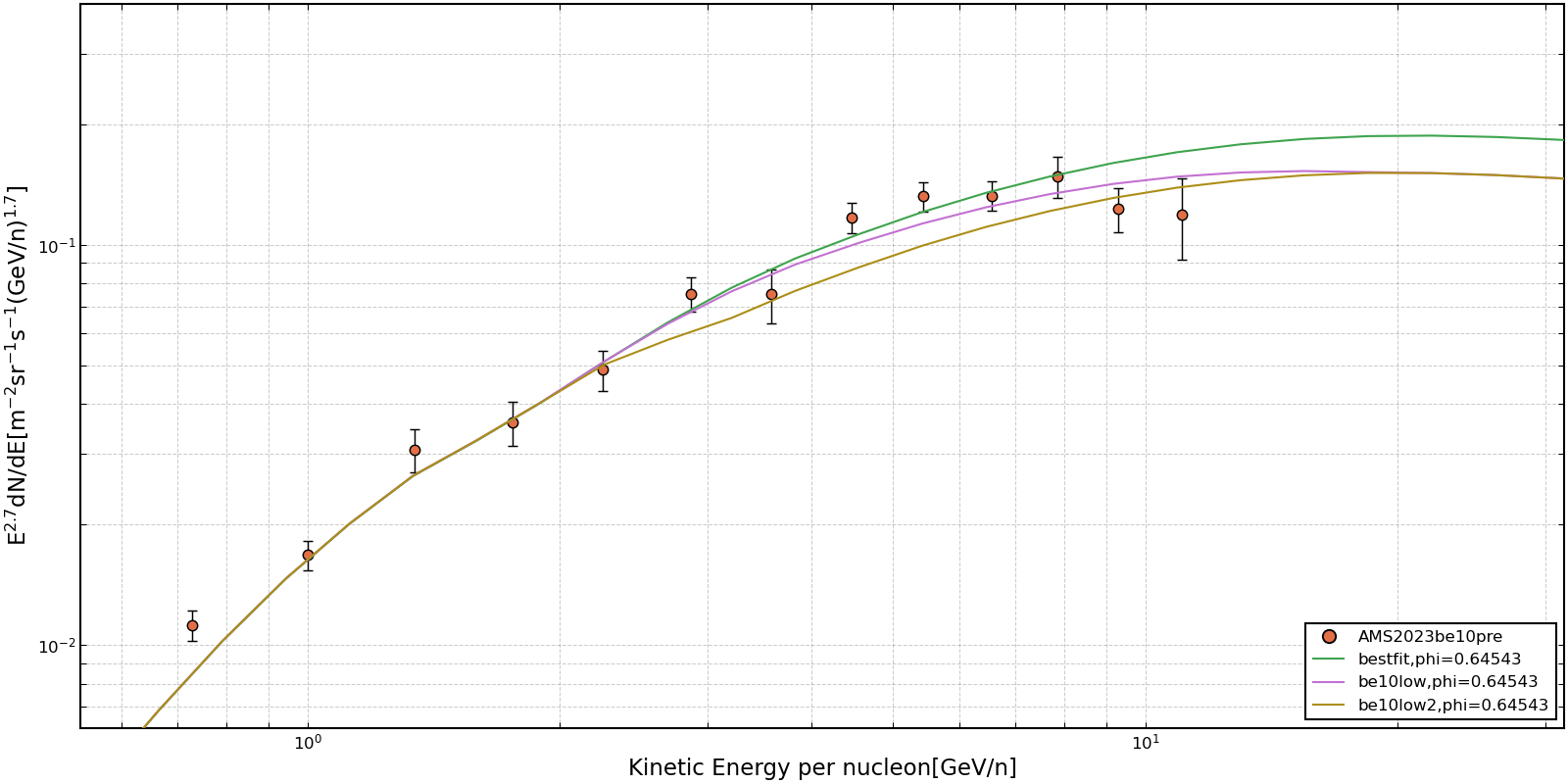}
\caption{\label{fig:BE10diff} 
Top panel: Cross-section measurements of the channel $\rm O + p \longrightarrow ^{10}\text{Be}$, compared with three possible parametrizations based on different adoption of data.
Bottom panel: The calculated of $\rm^{10}\text{Be}$ fluxes based on three possible parametrizations.}
\end{figure}

The extrapolation of $\rm^{10}\text{Be}$ fluxes is still uncertain since the cross-section data points above 3 GeV/$n$ exhibit an unexpected suppression. To analyze the impact of the suppression, we assume three parametrizations that determine the position of the plateau differently, named after \textit{bestfit}, \textit{be10low}, and \textit{be10low2}. The \textit{bestfit} parametrization is obtained without fitting these data points above 3 GeV/$n$, considering that they are significantly ($\sim~70\%$) lower than the energy-independent expectation.
The calculated fluxes are shown in Fig.~\ref{fig:BE10diff}. The \textit{be10low2} parametrization (yellow line) significantly under-predicts the $\rm^{10}\text{Be}$ fluxes compared with the AMS-02 observation and is thus disfavored.
The cross-section measurement at 3.25 GeV/$n$ from [Ba05] \cite{Bazarov:2005qu} may not be accurate, which is also disapproved by the above analysis of the B production channels.
The green line illustrates the best-fit result of the parametrization we used in the combined fitting, and the purple line illustrates the \textit{be10low} parametrization that considered the last point from [Yi69] \cite{PhysRev.166.968}. The latter allows a slower decline, and the cross section becomes constant at above 19 GeV/$n$. The predicted $\rm^{10}\text{Be}$ flux is a bit lower than the best-fit result but shows a decreasing trend that might be implied from the final two points of AMS-02 measurement \cite{AMS:2023anq}. Thus, the \textit{bestfit} and \textit{be10low} parametrizations are both acceptable in terms of interpreting the $^{10}$Be flux. 

\begin{figure}[htbp]
\includegraphics[width=0.5\textwidth,trim=0 0 0 0,clip]{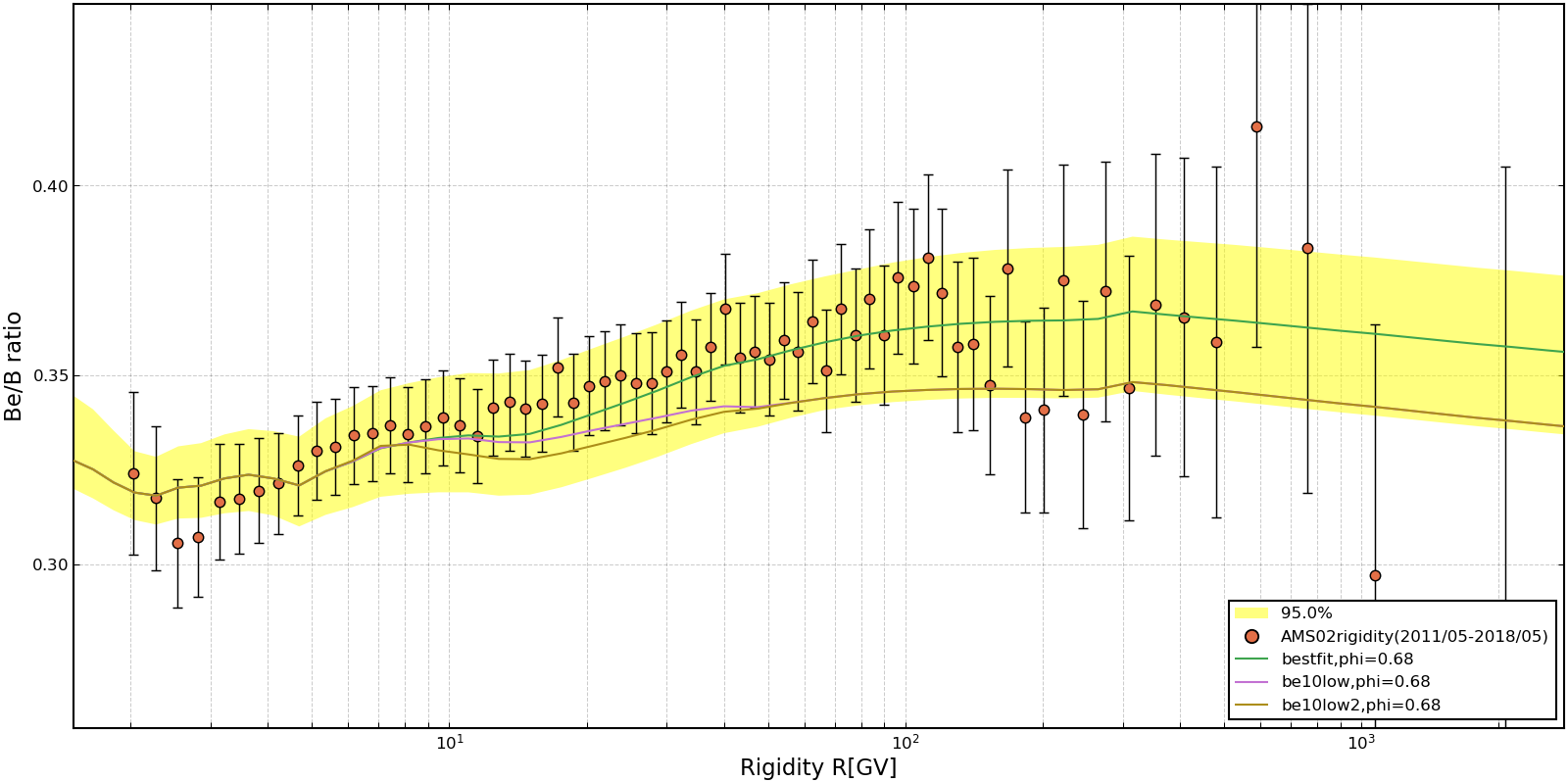}
\caption{\label{fig:BEBdiff} 
The prediction of Be/B ratios based on three possible parametrizations defined in the Fig.~\ref{fig:BE10diff} and the 2$\sigma$ confidence interval (yellow band) of the \textit{bestfit} parametrization, compared with the AMS-02 observation \cite{Aguilar:2021tos}.}
\end{figure}

To distinguish the extrapolation of the $\rm^{10}\text{Be}$ cross-section parametrization, we use the Be/B ratio provided by AMS-02 \cite{Aguilar:2021tos} as a reference, which is measured up to $\sim1$~TV. We simply assume that the cross section of $\rm^{9}\text{Be}$ follows the parametrization introduced in Appendix~\ref{app:xsdata}, which becomes constant at above 1 GeV/$n$ that is preferred by the $\rm^{9}\text{Be}$ flux (see Sec.~\ref{sec:be/b}).
In Fig.~\ref{fig:BEBdiff}, the predictions of the three parametrizations are compared with the measurements.
The highest result (green line) fits the AMS-02 data best, while those with much smaller cross section may be disfavoured as their overall Be fluxes are not sufficient to reproduce the Be/B ratio.
Since the total Be flux is determined by the isotope $\rm^{7}\text{Be}$, $\rm^{9}\text{Be}$, and $\rm^{10}\text{Be}$ fluxes, it is necessary to analyze the $\rm^{9}\text{Be}$ production, to determine a proper extrapolation of the Be/B ratio. In the next section, we will show that the parametrizations of \textit{bestfit} and \textit{be10low} can hardly be distinguished, considering the unclear determination of the $\rm^{9}\text{Be}$ cross section at high energies. Nonetheless, the diffusion halo size is mainly determined by the $\rm^{10}\text{Be}$ flux at $\leq10$ GeV/$n$, which would not be affected by the assumption of the cross-section extrapolation in higher-energy range.

\section{Lessons from the AMS-02 measurements of Be-9 and Be/B\label{sec:be/b}}
The uncertainty of $\rm^{9}\text{Be}$ cross-section measurements is much larger than that of $^{7}\text{Be}$, and we did not use the $\rm^{9}\text{Be}$ observation from AMS-02 \cite{AMS_ichep} to constrain the model parameters in Sec.~\ref{sec:result}.
In this section, we conversely use the precise $\rm^{9}\text{Be}$ flux to give a constraint to its cross-section production.

As illustrated in Appendix~\ref{app:xsdata}, the $\rm^{9}\text{Be}$ cross-section uncertainty is dominated by the $\rm O + p \longrightarrow ^{9}\text{Be}$ channel.
To transfer the uncertainty of the $\rm^{9}\text{Be}$ flux to the uncertainty of this dominant cross-section channel, we adopt the routine:
\begin{enumerate}
    \item For each energy bin [$E_i$,$E_i+\Delta E_i$] provided by the AMS-02 measurement \cite{AMS_ichep}, we calculate the renormalization factor $k_i=y_i^{\rm data}/y_i^{\rm model}$ and its uncertainty $r_i=\sigma_i^{\rm data}/y_i^{\rm model}$, where $y_i^{\rm model}$ is the $\rm^{9}\text{Be}$ flux calculated with the best-fit propagation parameters determined in Sec.~\ref{sec:result}
    \item To transfer the total $\rm^{9}\text{Be}$ flux to the contribution of only the $\rm O + p \longrightarrow ^{9}\text{Be}$ channel, we calculate the fraction of the reaction \cite{Genolini:2018ekk} with 
\begin{equation}
	f_{abc}=\frac{\psi-\psi(\sigma^{a+b\rightarrow c}=0)}{\psi}\,.
\end{equation}
The renormalization factor $k_0$ for specific channel can be calculated with $f_{abc}*(k_0-1)=k-1$, and its error $f_{abc}*r_0=r$.
   \item As the observed data corresponded to the post-modulated $\rm^{9}\text{Be}$ flux, the original energy bin shall be [$E_i$,$E_i+\Delta E_i$]+$\phi\frac{Z}{A}$, where $\phi=0.645$ GV, Z=4, A=9.
   \item The renormalized cross section for each energy bin is calculated with $\sigma=\sigma_0*(k_0\pm r_0)$, where $\sigma_0$ is the parametrization we adopted in the work, which is equal to 2.4 mb above 1 GeV/$n$.
\end{enumerate}

\begin{figure}[htbp]
\includegraphics[width=0.5\textwidth]{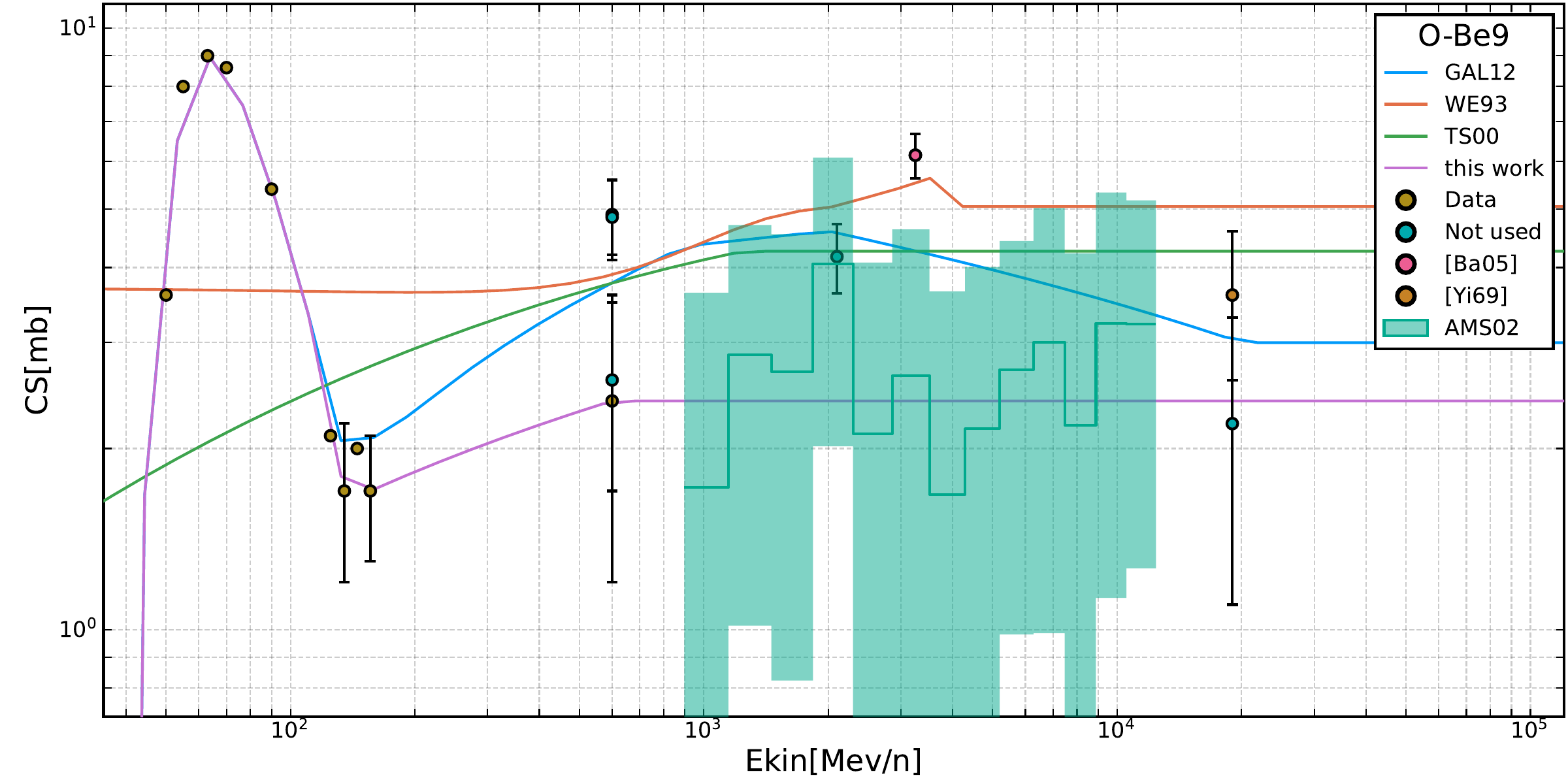}\\
\caption{\label{fig:xsbe9}
Cross section of $\rm O + p \longrightarrow ^{9}\text{Be}$ constrained by the AMS-02 $^9$Be data (green band), compared with the available cross-section measurements by different experiments. The reference parametrizations are taken from the {\footnotesize GALPROP} code.
}
\end{figure}

In Fig.~\ref{fig:xsbe9}, we illustrate the calculated result of the uncertainty band\footnote{The expectation should be wider if we consider the confidence interval of all parameters, but narrower if consider the systematical data connection between different energy bins and also the uncertainties from other reaction channels. Here we simply ignored those effects.} for the $\rm O + p \longrightarrow ^{9}\text{Be}$ channel. The AMS-02 observation prefers a much lower cross section, which may disfavor some cross-section measurements.
The highest and most disfavoured one is the [Ba05] experiments, which measured the cross section at 3.25 GeV/$n$.
In the figure, the default parametrization [GAL12] and the parametrizations of [WE93] (WNEW) and [TS00] (YIELDX) taken from the {\footnotesize GALPROP} code seem to systematically overpredict the cross section, implying that a significant reduction should be applied to better reconstruct the $\rm^{9}\text{Be}$ flux.
Above 12 GeV/$n$, the cross section is not constrained due to the maximum observed energy of the AMS-02 isotope. The cross section measurements at 19 GeV/$n$ are from [Yi69] \cite{PhysRev.166.968} and [RV84] \cite{READ1984359}, making the extrapolation result of $\rm^{9}\text{Be}$ unable to be distinguished yet.

We notice that the [Yi69] \cite{PhysRev.166.968} data gives unexpected cross section observation at high energies. In Fig.~\ref{fig:BE10diff}, we analyzed that the inclusion of [Yi69] can predict a decline of cross section, and the extrapolation of $\rm^{10}\text{Be}$ flux is lower. While in Fig.~\ref{fig:xsbe9} the inclusion of [Yi69] can predict an increment of cross section, making the extrapolation of $\rm^{9}\text{Be}$ flux larger. Combining these assumptions, we can obtain new results that are different from what we have got without using [Yi69] data. 

\begin{figure}[htbp]
\includegraphics[width=0.49\textwidth,trim=3 0 0 10,clip]{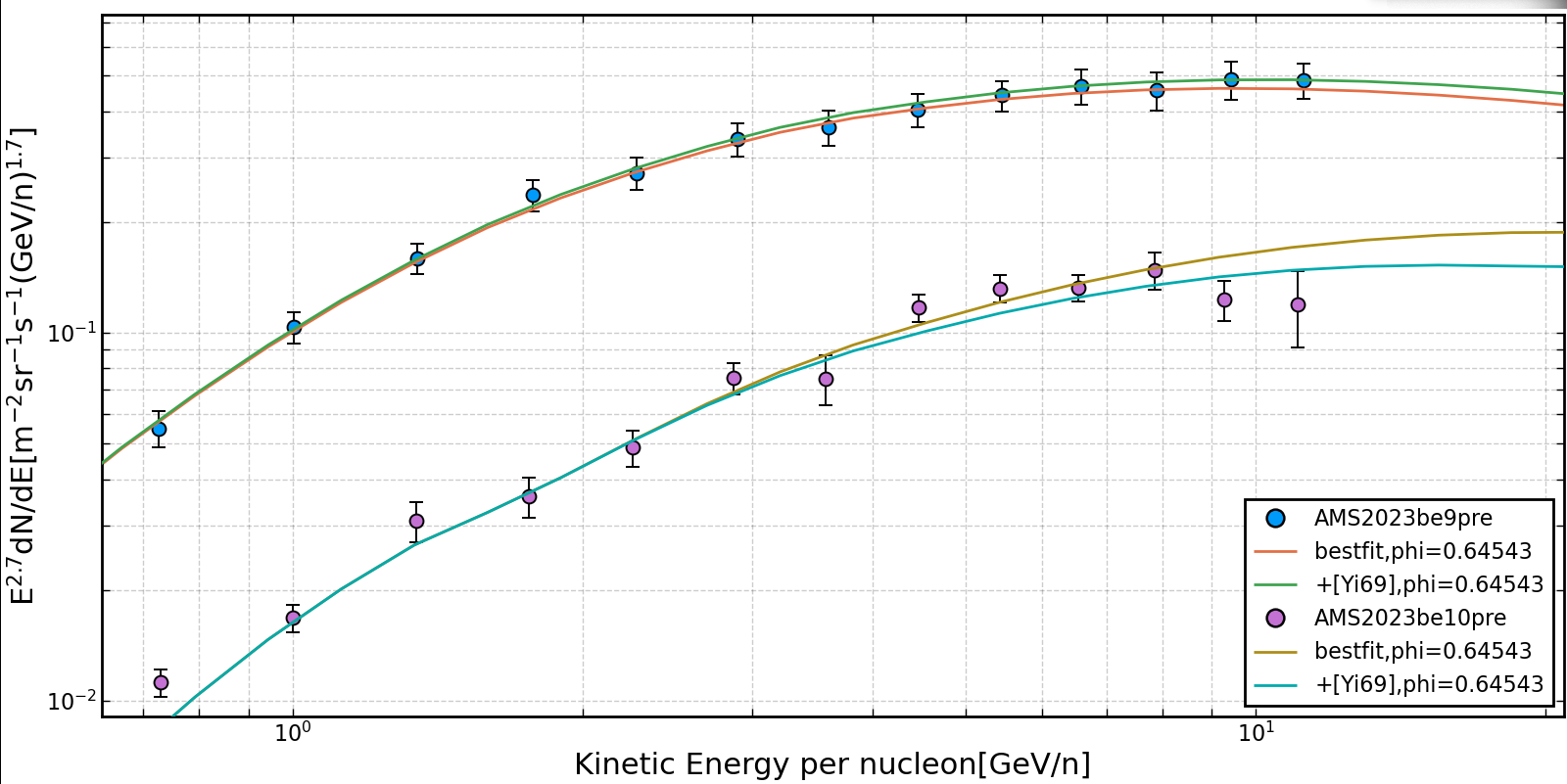}
\includegraphics[width=0.49\textwidth]{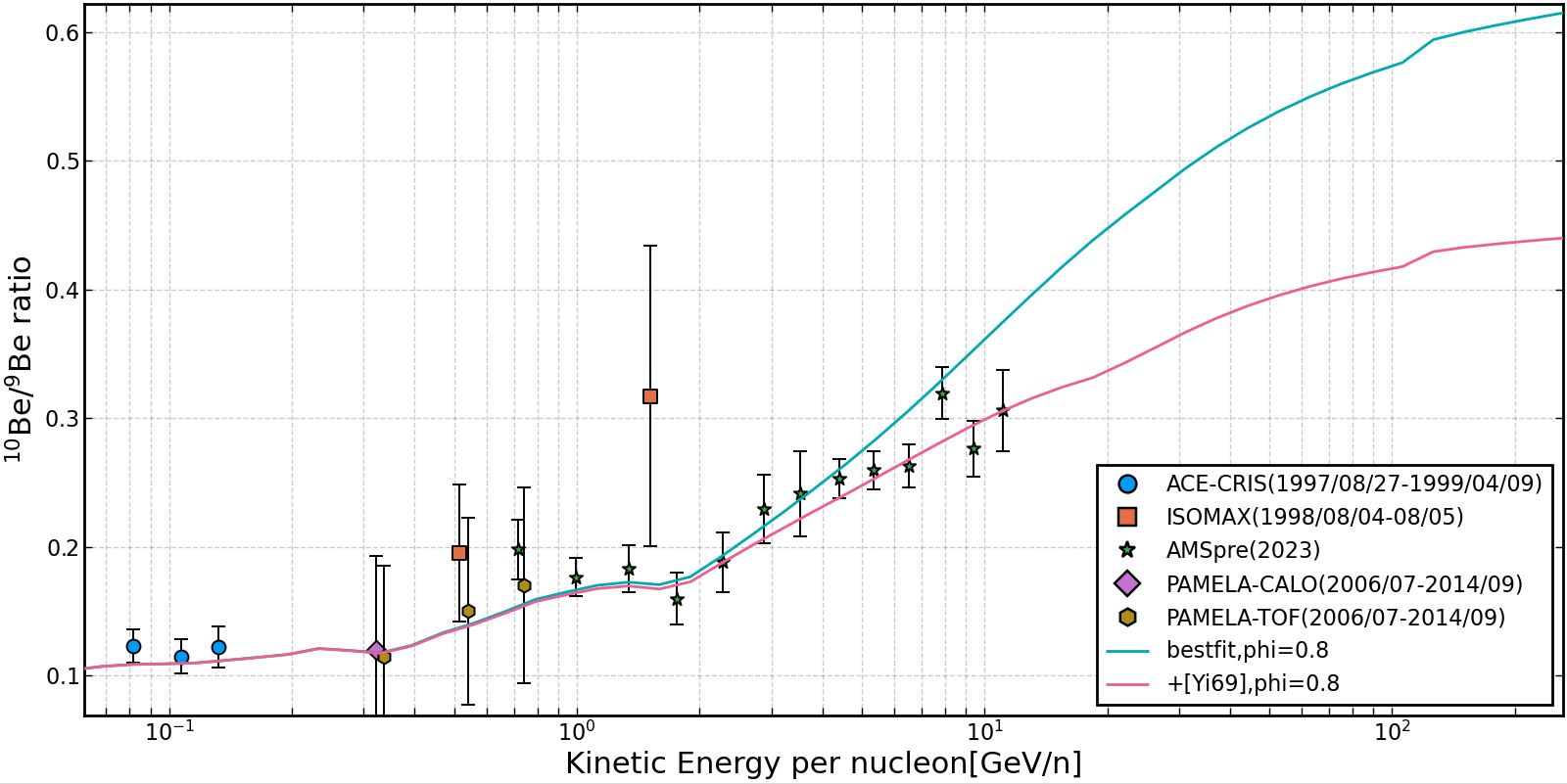}
\includegraphics[width=0.49\textwidth,trim=2 0 0 0,clip]{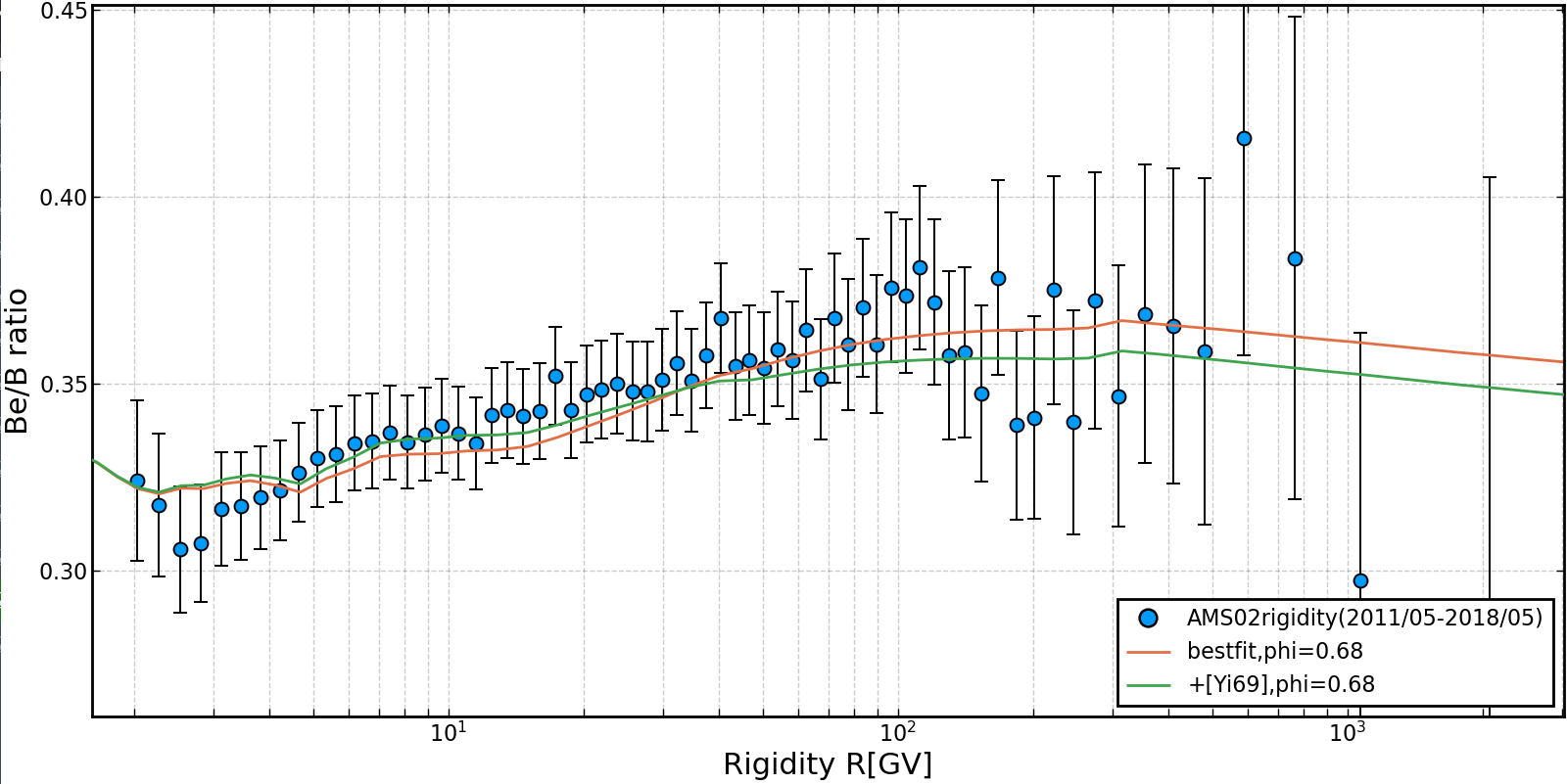}
\caption{\label{fig:yi69}Spectra calculated with or without [Yi69] data. Top panel: The $\rm^{9}\text{Be}$ and $^{10}\text{Be}$ spectra compared with AMS-02 measurements \cite{AMS_ichep}. Middle panel: The $\rm^{10}\text{Be}/^{9}\text{Be}$ ratio compared with measurements \cite{AMS_ichep,Hams:2004rz,2001ApJ...563..768Y,Nozzoli:2021vae}. Bottom panel: The Be/B ratio compared with AMS-02 measurements \cite{Aguilar:2021tos}.}
\end{figure}

In Fig.~\ref{fig:yi69}, we illustrate the $\rm^{9}\text{Be}$, $\rm^{10}\text{Be}$, $\rm^{10}\text{Be}/^{9}\text{Be}$, and Be/B result with or without using the [Yi69] data to predict the extrapolation of cross section. Generally, these two assumptions fit all the available observations well. 
In the top panel, the prediction of $\rm^{9}\text{Be}$ flux is higher when adopting the [Yi69] data, but still within the data constraints. The preliminary $\rm^{10}\text{Be}$ flux measured by AMS-02 implies an unexpected decline feature at above 10 GeV/$n$, which remains to be confirmed by higher-energy measurements. 
By including the [Yi69] data, we assume that the cross section at above 2 GeV/$n$ decrease steadily for $\rm^{16}\text{O} \longrightarrow ^{10}\text{Be}$ channel, whose validity requires more observations to prove. As a result, the predicted  $\rm^{10}\text{Be}$ flux follows a decline suggested by observation.
For almost the same reason, as shown in the middle panel, the prediction of $\rm^{10}\text{Be}/^{9}\text{Be}$ ratio at about 10 GeV/$n$ is better when using the [Yi69] data; the assumption of not using [Yi69] data predicts higher $\rm^{10}\text{Be}/^{9}\text{Be}$ ratio as the energy increases.
The best-fit result of $\rm^{10}\text{Be}/^{9}\text{Be}$ ratio predicts a bump structure at around 1 GeV/$n$, which has been discovered by the preliminary AMS-02 measurement. This structure is related to the $\rm^{10}\text{Be}$ cross-section bump shown in Fig.~\ref{fig:BE10diff}, which is constrained by a series of data from Michel's group \cite{Michel:1995xx, NSR1997MI26}.

As shown in Appendix \ref{app:xsdata}, most of the $\rm^{7}\text{Be}$ production cross section channels are strictly constrained by observations, while the contributions from $\rm^{9}\text{Be}$ and $\rm^{10}\text{Be}$ depend on how we extrapolate the cross sections. The AMS-02 Be/B ratio \cite{Aguilar:2021tos} may constrain the total Be flux at high energy and testify the assumption of $\rm^{9}\text{Be}$ and $\rm^{10}\text{Be}$. 
In the bottom panel of Fig.~\ref{fig:yi69}, we predict the Be/B ratio with or without using the [Yi69] data, but the results are so close that cannot be distinguished. The former result (green line) under-predicts the Be/B ratio at above 100 GV, while better reproducing the flux at about 10 GV compared with the latter result. As we have adjusted the cross section of B and Be at low energies according to the available cross-section measurements, the predicted Be/B ratio at several GVs shows similar trends as what AMS-02 measured.

It is important to decide the $\rm^{10}Be$ cross section since the production of $\rm^{10}Be$ directly impacts the determination of the halo height $L$. In Table~\ref{tab:best} the best-fit value of $L$ is 5.674 kpc, under the assumption that the $\rm^{16}\text{O} \longrightarrow ^{10}\text{Be}$ cross section is energy-independent above 3 GeV/$n$. If using the [Yi69] data, we find that the best-fit value of $L$ would decrease to $\sim 5$ kpc. The difference is less significant since the Be isotope fluxes remain unchanged at low energies where the halo height is more sensitive to the flux.

From a global analysis of the CR and cross-section data mentioned above, we prefer discarding the [Ba05] \cite{Bazarov:2005qu} data for channels of $\rm^{16}\text{O} \longrightarrow ^{11}\text{B}$, $\rm^{16}\text{O} \longrightarrow ^{11}\text{C}$, $^{16}\text{O} \longrightarrow ^{9}\text{Be}$ and $\rm^{16}\text{O} \longrightarrow ^{10}\text{Be}$. In Appendix \ref{app:xsdata}, the [KO99] and [KO02] \cite{Korejwo:2000pf, Korejwo:2002ts} data of several channels were ignored as they significantly deviate from other measurements.
Coincidentally, both groups (Bazarov et al. and Korejwo et al.) performed their experiments by using the synchrophasotron's beam at the Joint Institute for Nuclear Research in Dubna. 

We have also calculated the total Li flux and its isotopic composition with the best-fit parameters in Table~\ref{tab:best}. Compared with the AMS-02 measurements \cite{Aguilar:2021tos, AMS_icrc2021}, we found a 12\% overprediction of Li flux, which was also mentioned by Maurin \cite{Maurin:2022irz}. Our result of the $\rm^6Li$ spectrum significantly overpredicts the flux while the $\rm^7Li$ spectrum fits the preliminary AMS-02 measurements \cite{AMS_icrc2021} well. Those features are mostly attributed to the over-assumed cross section of the $\rm^6Li$, as most of its production channels are poorly determined with seldom or no data. The next step of this work would be to calculate and analyze these secondary nuclei to make constraints on their cross sections. 

\section{SUMMARY\label{sec:conclusion}}
The preliminary results of the Be isotope flux measurements provided by AMS-02 have reached an unprecedented energy of 12 GeV/$n$, which is of significant importance for constraining the propagation parameters of Galactic CRs. 
Moreover, the precision of the AMS-02 measurements is now significantly better than that of the production cross sections for certain isotopes. In light of this, we combine the high-precision cross sections of selected isotopes with the AMS-02 CR measurements to refine the propagation model. This model, in conjunction with the AMS-02 measurements, has been employed to inversely estimate the less constrained production cross sections.
This method should also be meaningful for studying the forthcoming isotopic data of other nuclei.

We employ the {\footnotesize GALPROP} code to calculate CR propagation and conduct a comprehensive Bayesian analysis using a MCMC sampling algorithm to derive posterior distributions of the model parameters. We meticulously account for the uncertainties in the production cross sections of secondary particles to achieve more robust constraints on the parameters. Our approach incorporates a data-driven parametrization of cross sections, fully utilizing the wealth of available experimental data. We have enriched the isotopic cross-section database within {\footnotesize GALPROP} by including additional data from the EXFOR database, as well as the fresh high-energy B production cross sections from the NA61/SHINE experiment. Moreover, we adopt an energy-dependent form for the uncertainty of the parametrization, thereby reflecting the increased uncertainty at higher energies where the cross sections are less stringent.

In the Bayesian analysis, we use the CR flux measurements of $^7$Be, $^9$Be, B, C, and O. Owing to the high-quality measurements of the $^7$Be production cross section, we have innovatively employed $^7$Be in place of $^9$Be to constrain propagation parameters. The resulting parameters are largely consistent with previous studies: the diffusion coefficient at 1 GeV is $\approx5\times10^{28}$~cm$^2$~s$^{-1}$, and the diffusion index is 0.43, aligning more closely with the expectation of Kraichnan's turbulence theory. Notably, the thickness of the diffusion halo is constrained to $5.67\pm0.76$~kpc, representing a moderate value compared to previous analogous works. Both as mid-mass stable secondary particles, B and $^7$Be data are well-fitted simultaneously, yielding acceptable $\chi^2$ for both the global fit and the cross-section part. This suggests that the CR measurements and the production cross-section measurements of $^7$Be and B are self-consistent.

Using the well-constrained CR propagation model and the $^9$Be cross-section parametrizations embedded in {\footnotesize GALPROP}, we discover a significant overestimation of the $^9$Be CR fluxes in comparison with the AMS-02 measurement. In response to the demands of the CR data, we impose a constraint to the cross section of the $\rm^{16}\rm{O} \longrightarrow ^{9}\rm{Be}$ channel in the energy range of $\sim1$ to $10$~GeV/$n$. The result indicates that the production cross section ought to be significantly lower than previously thought. In particular, we observe that the cross section at $3.25$~GeV/$n$ measured by the [Ba05] experiment \cite{Bazarov:2005qu} is significantly elevated beyond our expectation and is therefore considered less credible.

Intriguingly, the [Ba05] measurement for $\rm^{16}\rm{O} \longrightarrow ^{10}\rm{Be}$ is similarly challenged by the $^{10}\rm{Be}$ spectrum of AMS-02. Furthermore, while the $^7\rm{Be}$ and B data can be explained consistently, the required renormalization factors for B production channels are slightly but systematically higher, which may not be ascribed to random fluctuations in cross-section observations. This discrepancy is resolved when the [Ba05] cross-section measurements for $\rm^{16}\text{O} \longrightarrow ^{11}\text{B}$ and $\rm^{16}\text{O} \longrightarrow ^{11}\text{C}$ channels are excluded. This could be the first instance of picking up unreliable nucleon production cross-section data through the CR study, rather than through inconsistencies among different cross-section measurements.

The possible high-energy extrapolations of the Be isotopic spectrum are discussed.
We notice that the cross-section data of [Yi69]  \cite{PhysRev.166.968} for $\rm^{16}\text{O} \longrightarrow ^{9}\text{Be}$ and $\rm^{16}\text{O} \longrightarrow ^{10}\text{Be}$ channels, which are not used in our default calculations, indicate energy-dependent cross sections up to 19 GeV/$n$. It could influence the high-energy behavior of the extrapolated spectra.
By adopting the [Yi69] observations, we find that the outcomes remain consistent with the CR data of Be isotopes and Be/B, making it challenging to distinguish from the default scenario.
However, it is noteworthy that the inclusion of [Yi69] data yields a better fit to the high-energy tail of the AMS-02 $\rm^{10}\text{Be}$ (or $\rm^{10}\text{Be}/^{9}\text{Be}$) data.
The forthcoming balloon-borne experiment HELIX \cite{Park:2021oic, Wakely:2023iaq} can provide precise $\rm^{10}\text{Be}$ and $\rm^{9}\text{Be}$ measurements in the energy range of $0.2\sim3$ GeV/$n$, and a new magnet spectrometer would be subsequently payload to measure up to 10 GeV/$n$. This may give a cross-check to the high-energy features observed by AMS-02, thereby offering a critical evaluation of the associated cross-section measurements.

 
\acknowledgments
This work is supported in part by the National Natural Science Foundation of China under Grants No. 12042507, No. 12105292, and No. 12175248. 

\bibliography{apssamp}

\appendix

\section{CROSS SECTION DATA\label{app:xsdata}}
Here we show the plots of the most important channels needed for analyzing Be and B production in the paper.
Secondary CRs arise from the fragmentation of heavier nuclei upon collision with the ISM gas, which is composed of mostly hydrogen and helium gas. In the section, we only present channels of collisions with the hydrogen target for simplicity. As implemented in {\footnotesize GALPROP}'s fragmentation routine, the collisions with the helium target are calculated using a parametrization by Ferrando  \cite{Ferrando:1988tw} where the interstellar gas ratio of helium to hydrogen is set to be 0.11.

Figures below show the comparison between the model and measurements for the relevant channels. The available data are obtained from:
\begin{enumerate}
    \item\texttt{isotope\_cs.dat}: The isotopic cross-section database file is built in the {\footnotesize GALPROP} code \cite{Moskalenko:2001qm,Moskalenko:2003kp} for normalizing the parametrization formulae, such as \texttt{WNEW} code by Webber \cite{Webber:1990kc,Webber:1998ex,Webber_2003} or \texttt{YIELDX} code by Tsao and Silberberg \cite{Silberberg:1998lxa}.
The cross section data assembled in the file were taken from multiple cross-section measurements published before 2003.

\item EXFOR (Experimental Nuclear Reaction Data): The website\footnote{\url{https://www-nds.iaea.org/exfor}.} is an extensive database containing experimental data, as well as bibliographic information, experimental setup, and source of uncertainties. By querying the EXFOR database, we can add most of the measurements published so far.

\item NA61/SHINE: Additional measurements are reported by the NA61/SHINE Collaboration \cite{Amin:2021oow, Amin:2023fki} in the International Cosmic Ray Conference (ICRC).
The recent pilot run provided precise high-energies measurements of cross sections from the C projectile at 13.5 GeV/$n$, which is valuable for constraining the uncertainties of the dominant channels.
\end{enumerate}

In this work, we analyzed a data-driven parametrization by using the default evaluation routine implemented in the {\footnotesize GALPROP} code (see \texttt{nuc\_package.cc} for details). The routine checks if there exists cross-section data for corresponding channels given in the file \texttt{eval\_iso\_cs.dat} and if so interpolates it linearly. Otherwise, the routine would use the parametrization formulae of \texttt{WNEW} or \texttt{YIELDX} and re-normalize it according to the data given in \texttt{isotope\_cs.dat}.
 Webber's parametrization doesn't define the secondary Li production, and only the contributions of the dominant channels are introduced in \texttt{eval\_iso\_cs.dat} for calculating the Li flux when adopting the [GAL12] parametrization. It has been pointed out \cite{Maurin:2022irz} that the Li flux predicted by [GAL12] is significantly lower than that of [GAL22], as the former doesn't properly calculate the Li production from Fe projectile channels and lacks other missing channels.
To improve that, we use the parametrization provided by Silberberg and Tsao instead for specific channels, when Webber's parametrization cannot provide a non-zero cross section.

\begin{figure*}[htbp]
\caption{\label{channel1}Channels: $\rm C + p \longrightarrow \text{B}$ or $\text{Be}$ isotopes. The parametrizations [GAL12] shown in the figures were taken from the {\footnotesize GALPROP} code.}
    \centering
      \subfigure{
        \includegraphics[width=0.45\textwidth]{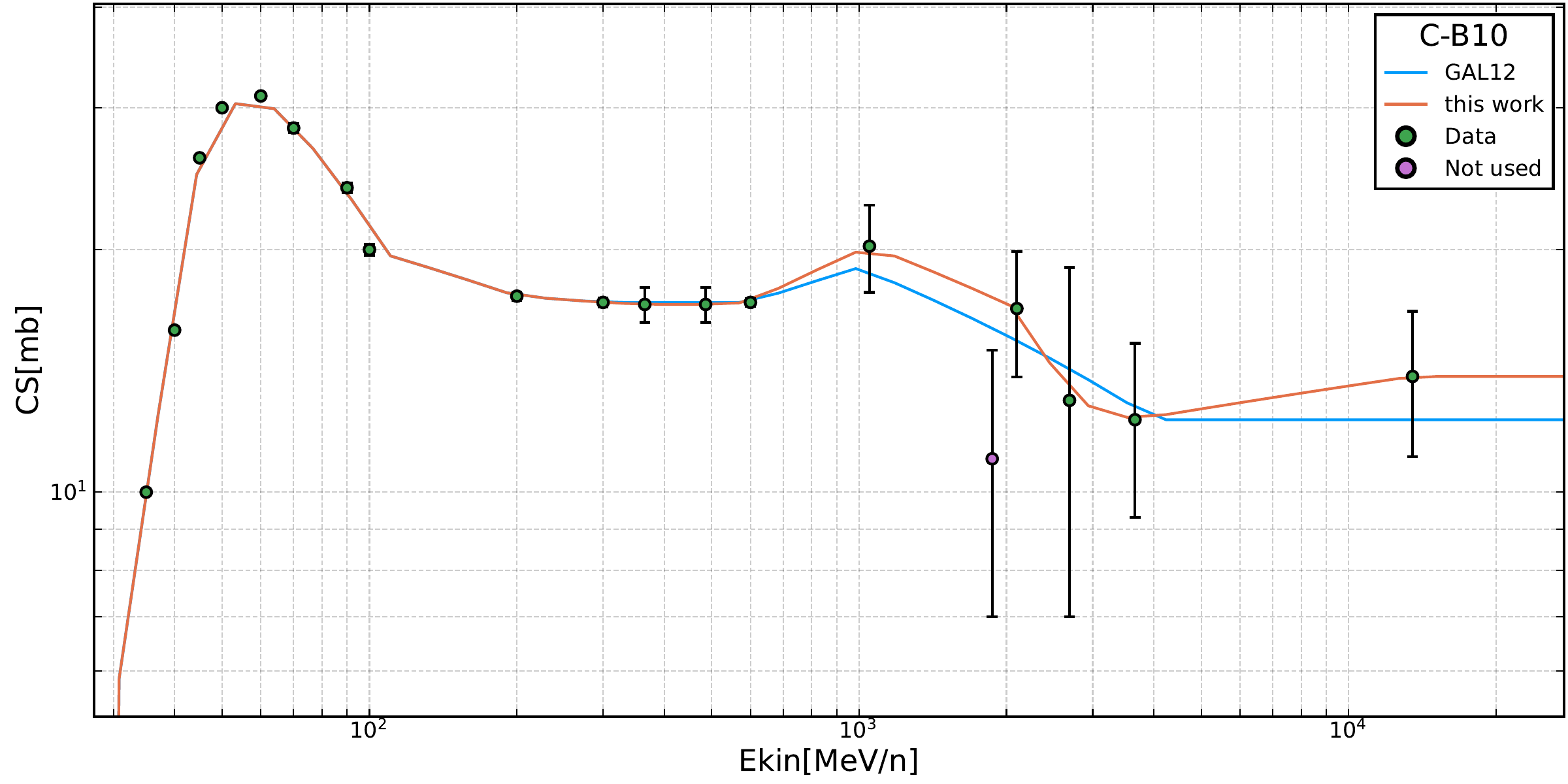}
      }
      \subfigure{
        \includegraphics[width=0.45\textwidth]{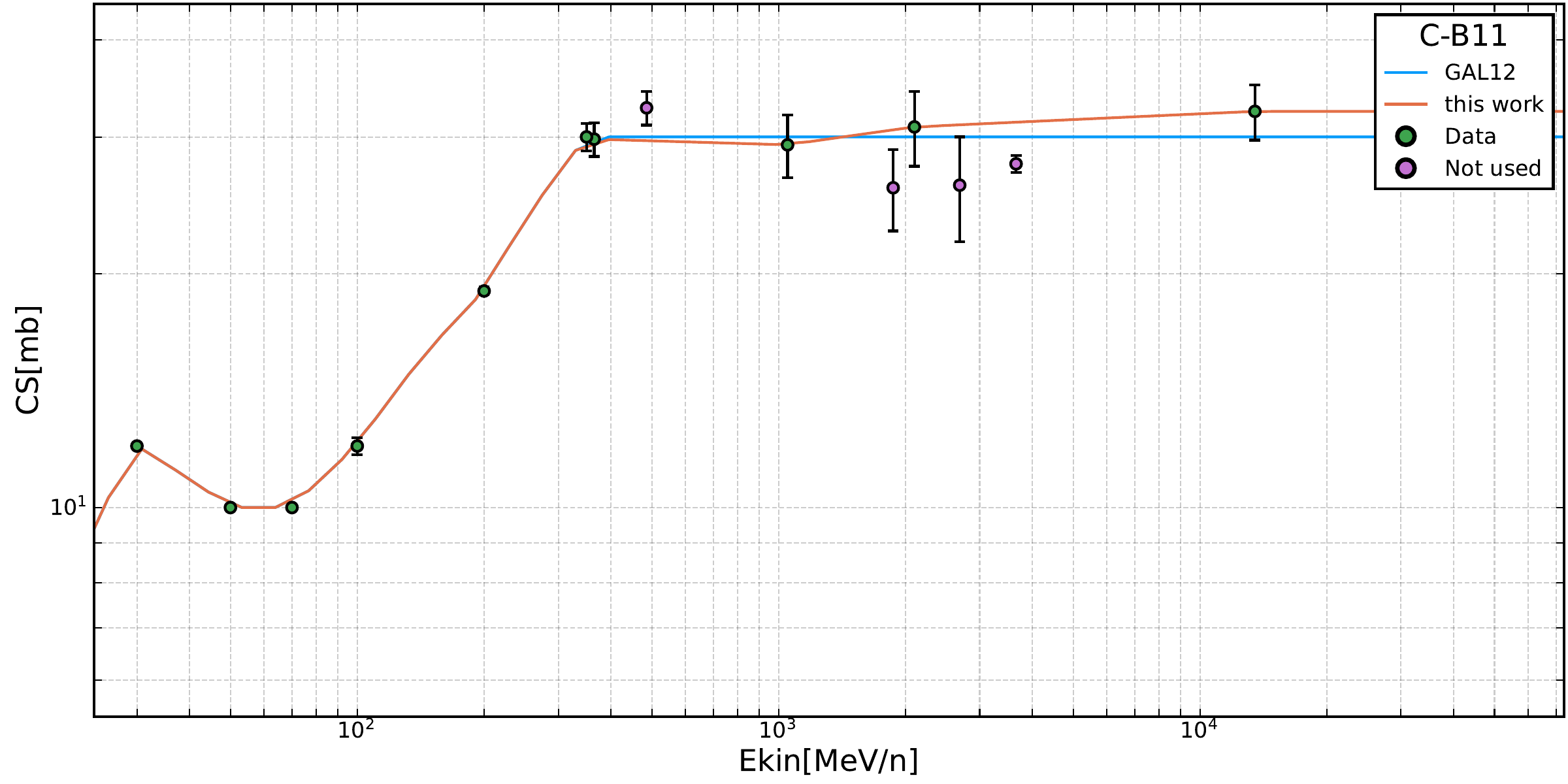}
      } \\
      \subfigure{
           \includegraphics[width=0.45\textwidth]{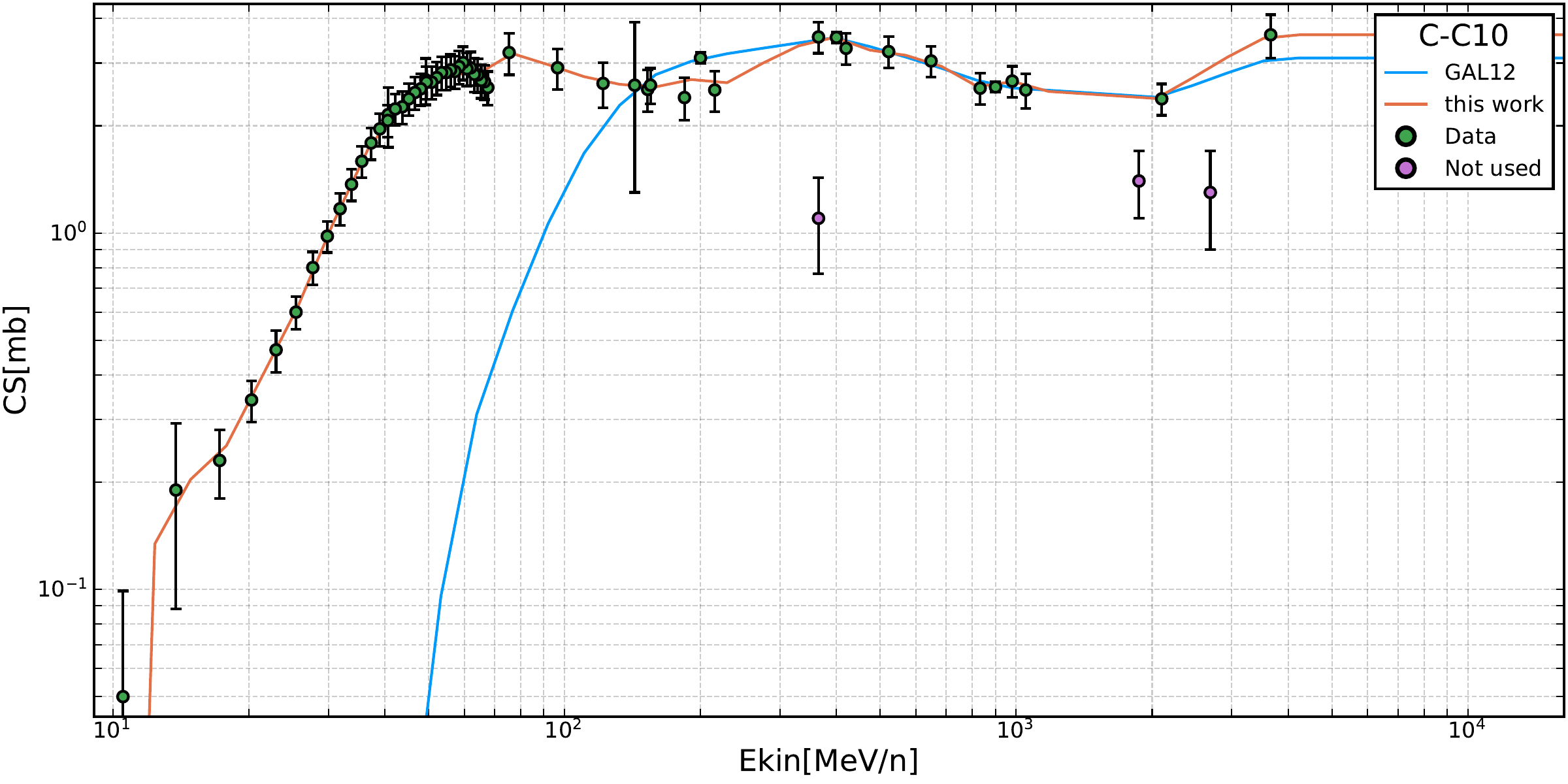}}
       \subfigure{    
           \includegraphics[width=0.45\textwidth]{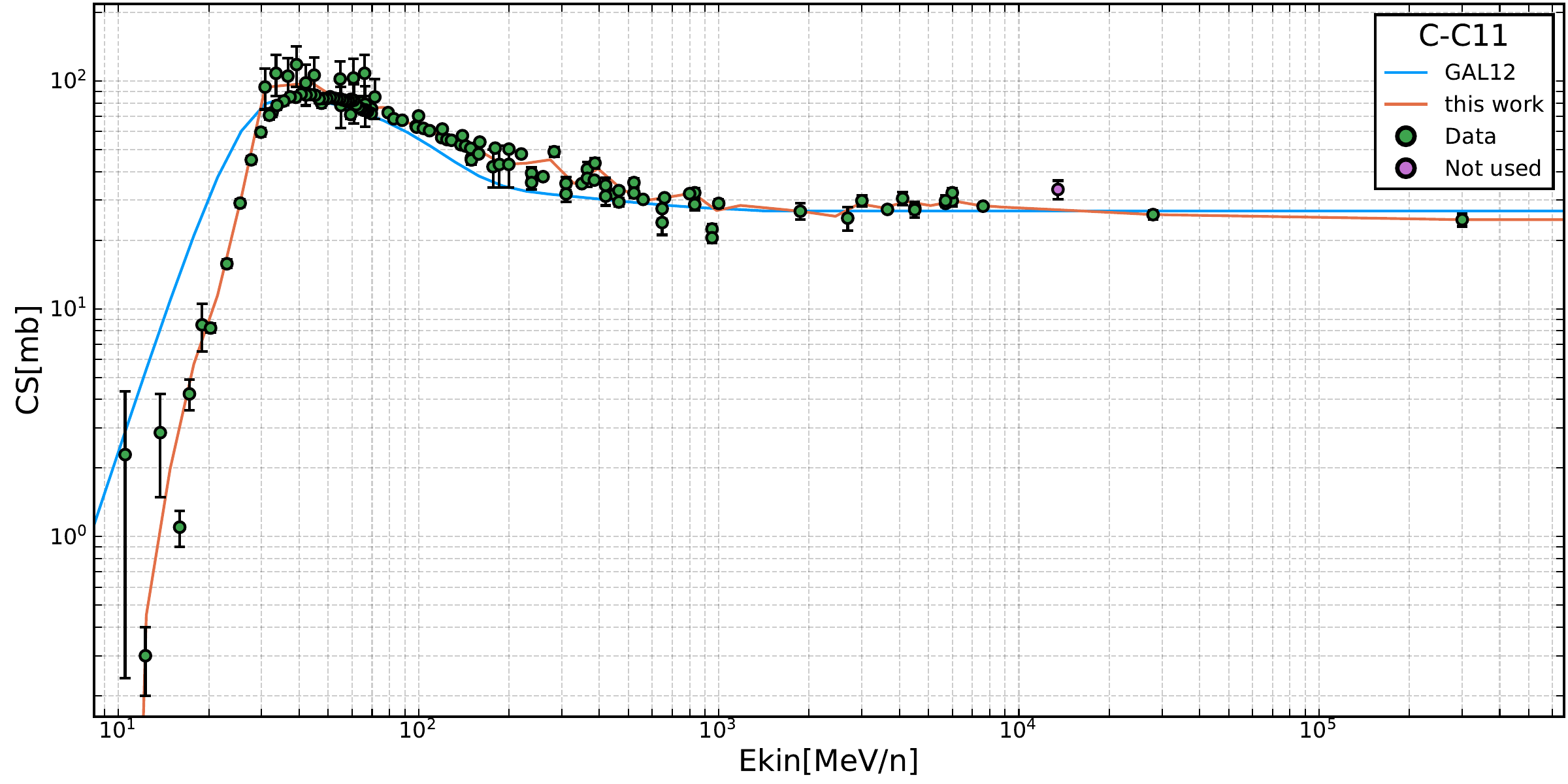}
      }\\
 
      \subfigure{
         \includegraphics[width=0.45\textwidth]{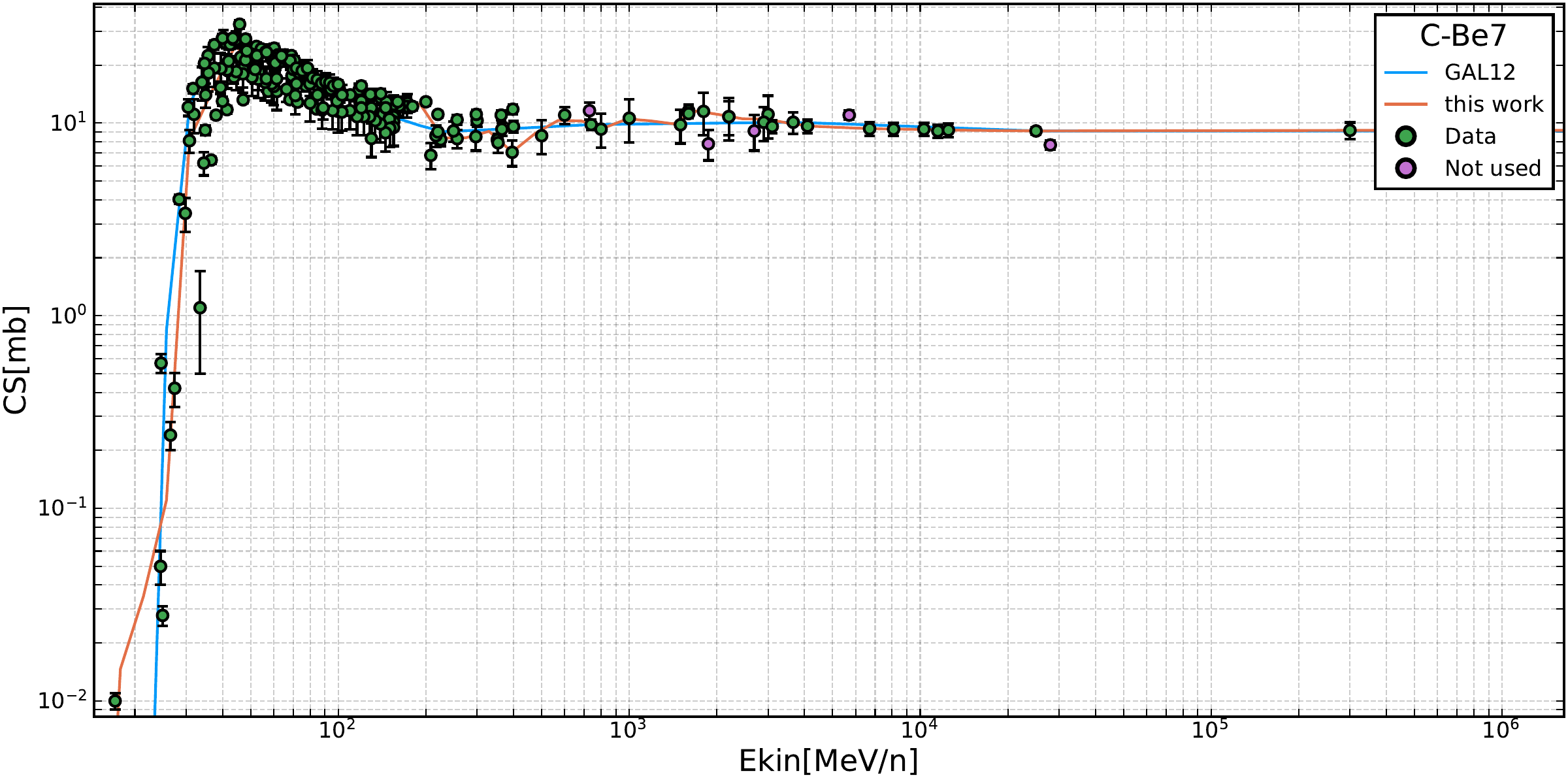}}
        \subfigure{  
           \includegraphics[width=0.45\textwidth]{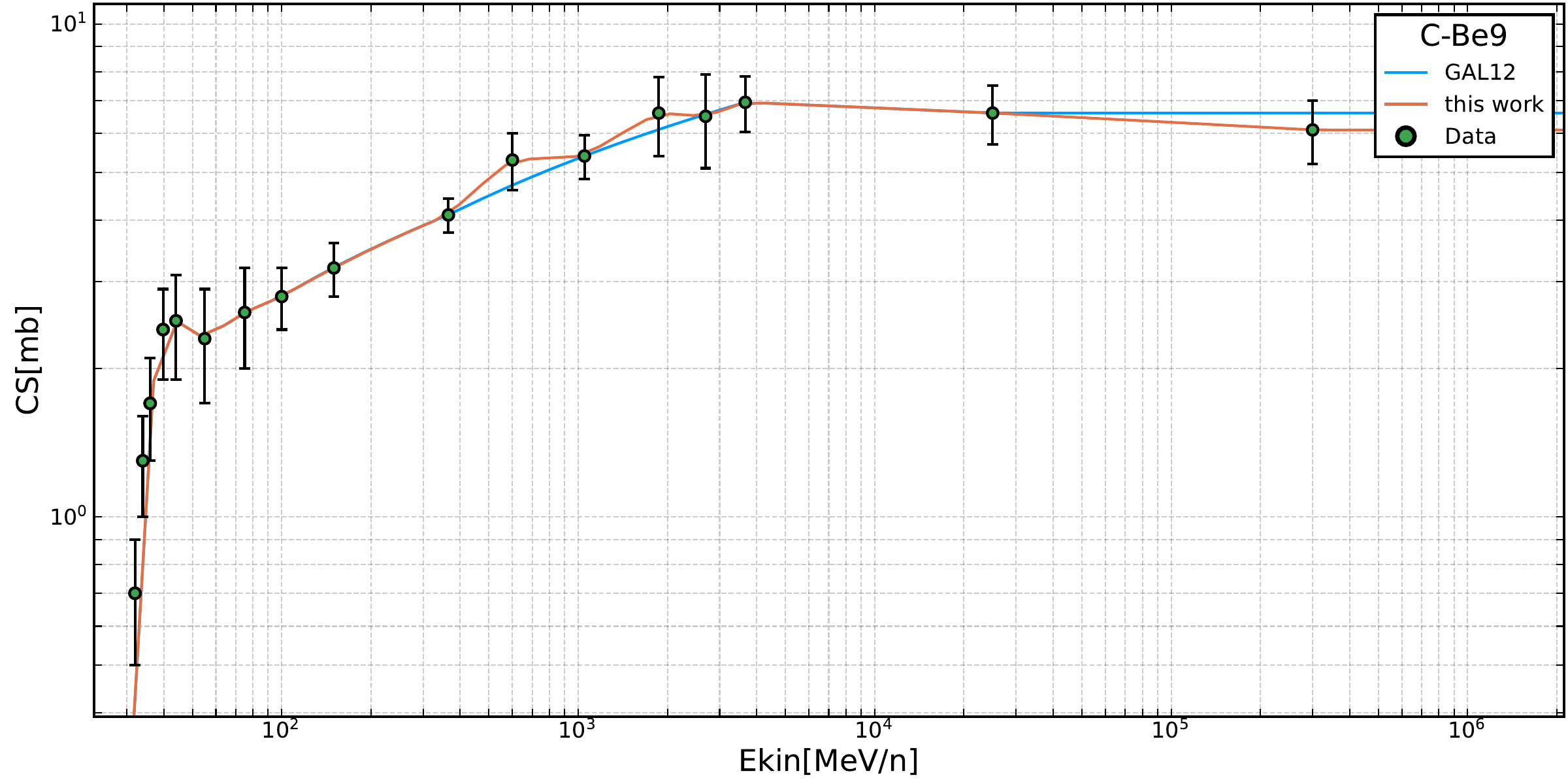}
      }\\
      \subfigure{
           \includegraphics[width=0.45\textwidth]{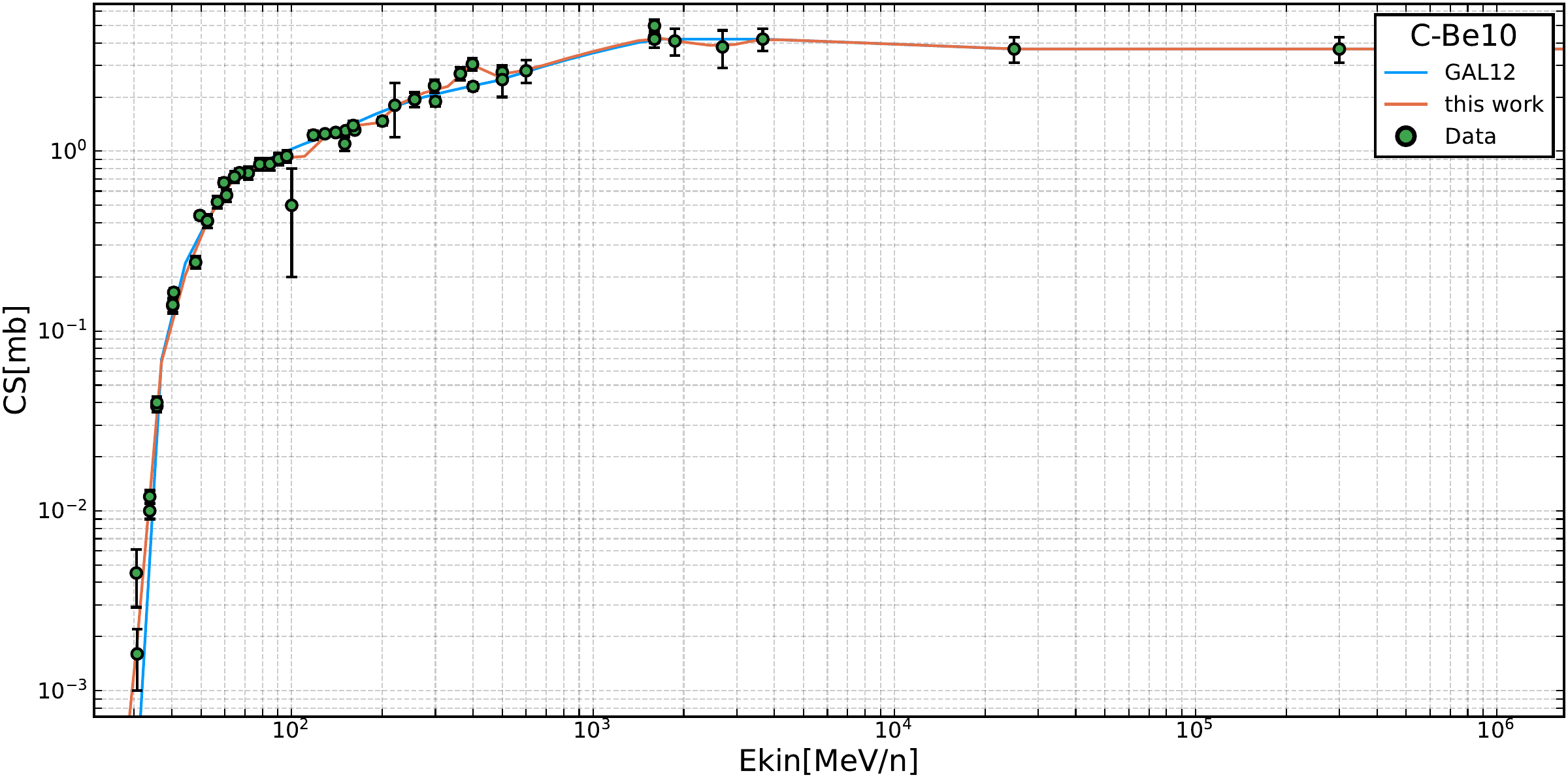}}
\end{figure*}

\begin{figure*}[htbp]
\caption{\label{channel2}Channels: $\rm O + p \longrightarrow \text{B}$ or $\text{Be}$ isotopes. The parametrizations [GAL12] shown in the figures were taken from the {\footnotesize GALPROP} code.}
    \centering
      \subfigure{
        \includegraphics[width=0.45\textwidth]{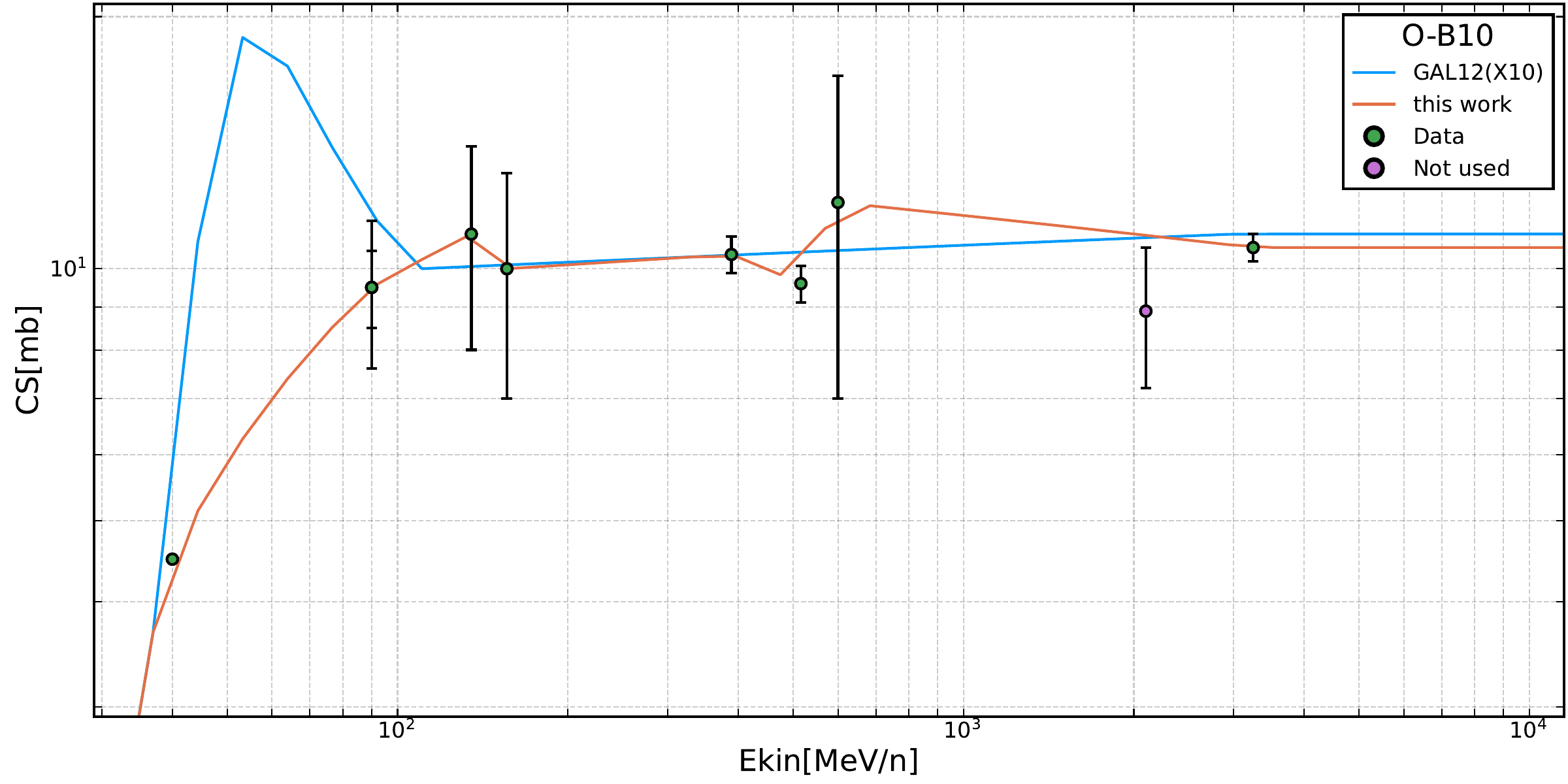}
      }
      \subfigure{
        \includegraphics[width=0.45\textwidth]{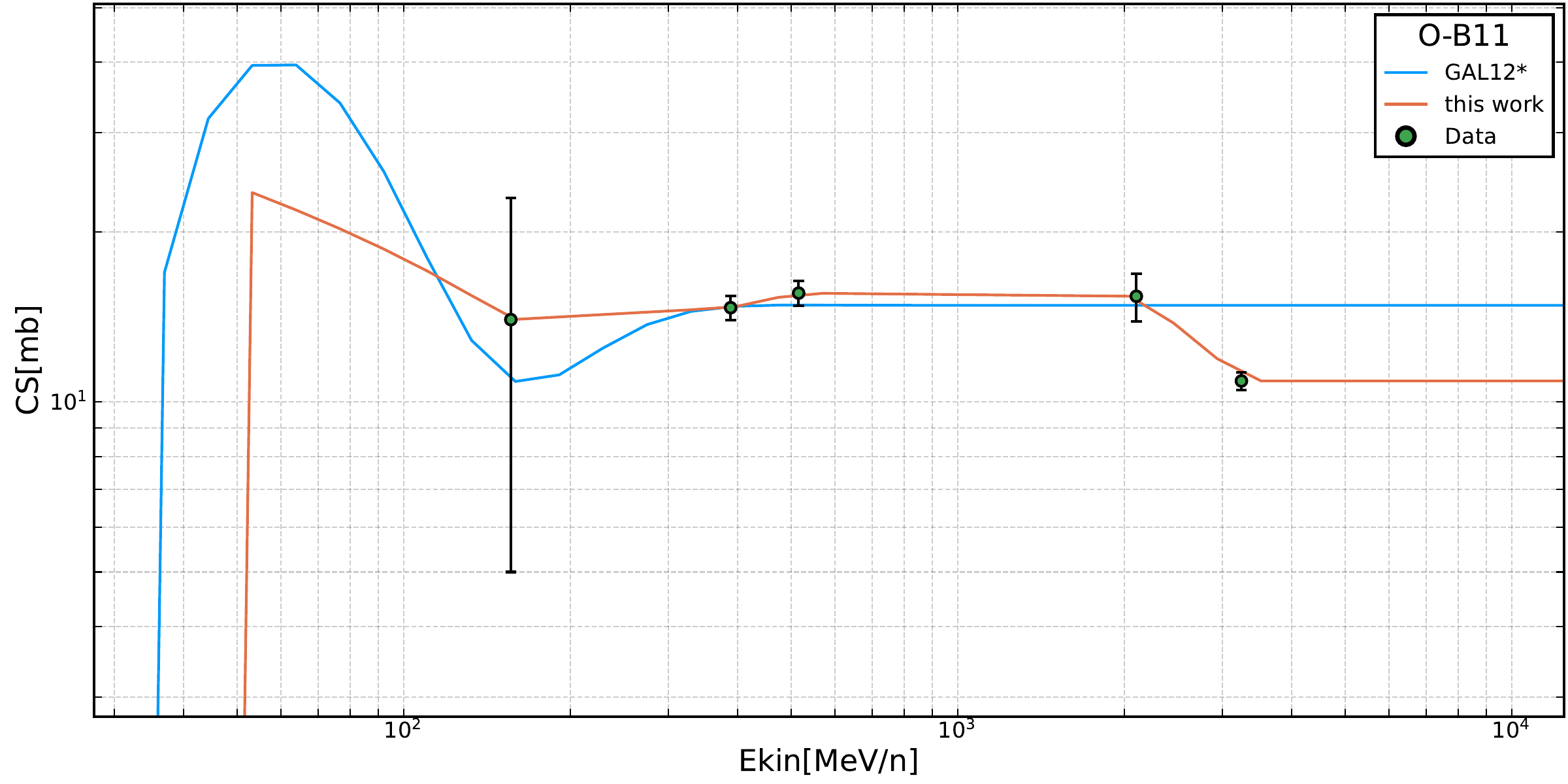}
      } \\
      \subfigure{
           \includegraphics[width=0.45\textwidth]{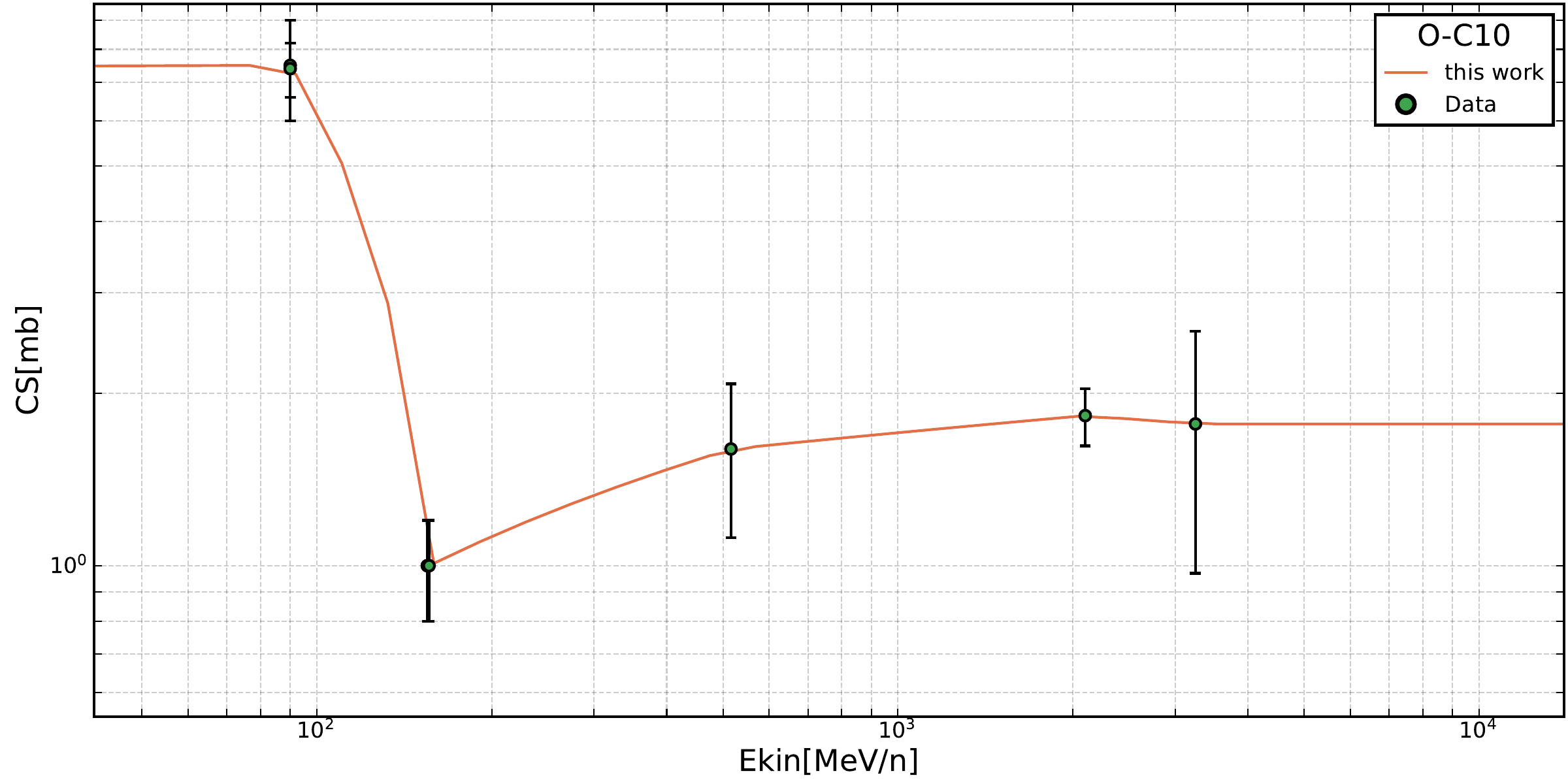}}
       \subfigure{    
           \includegraphics[width=0.45\textwidth]{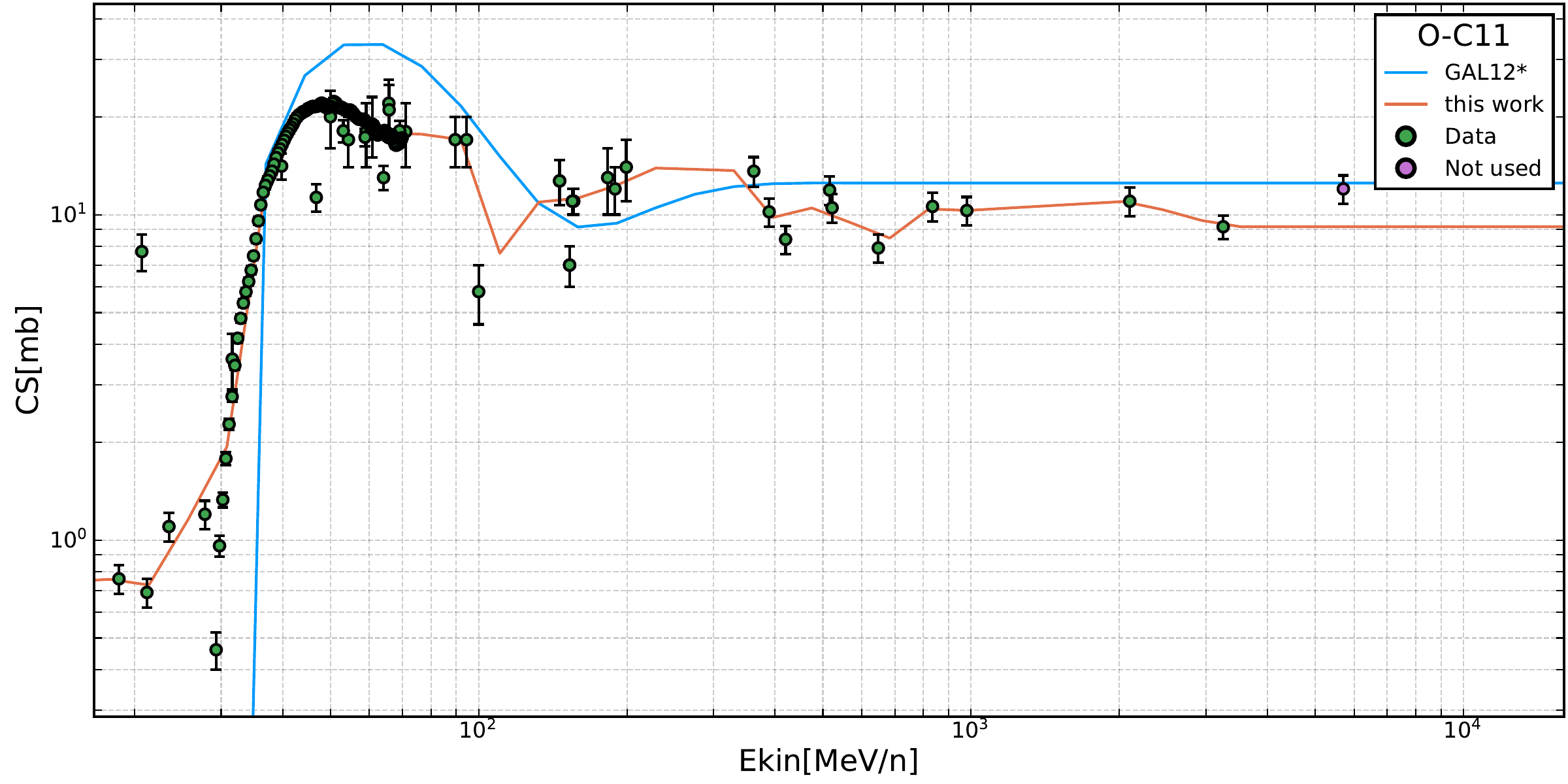}
      }\\
 
      \subfigure{
         \includegraphics[width=0.45\textwidth]{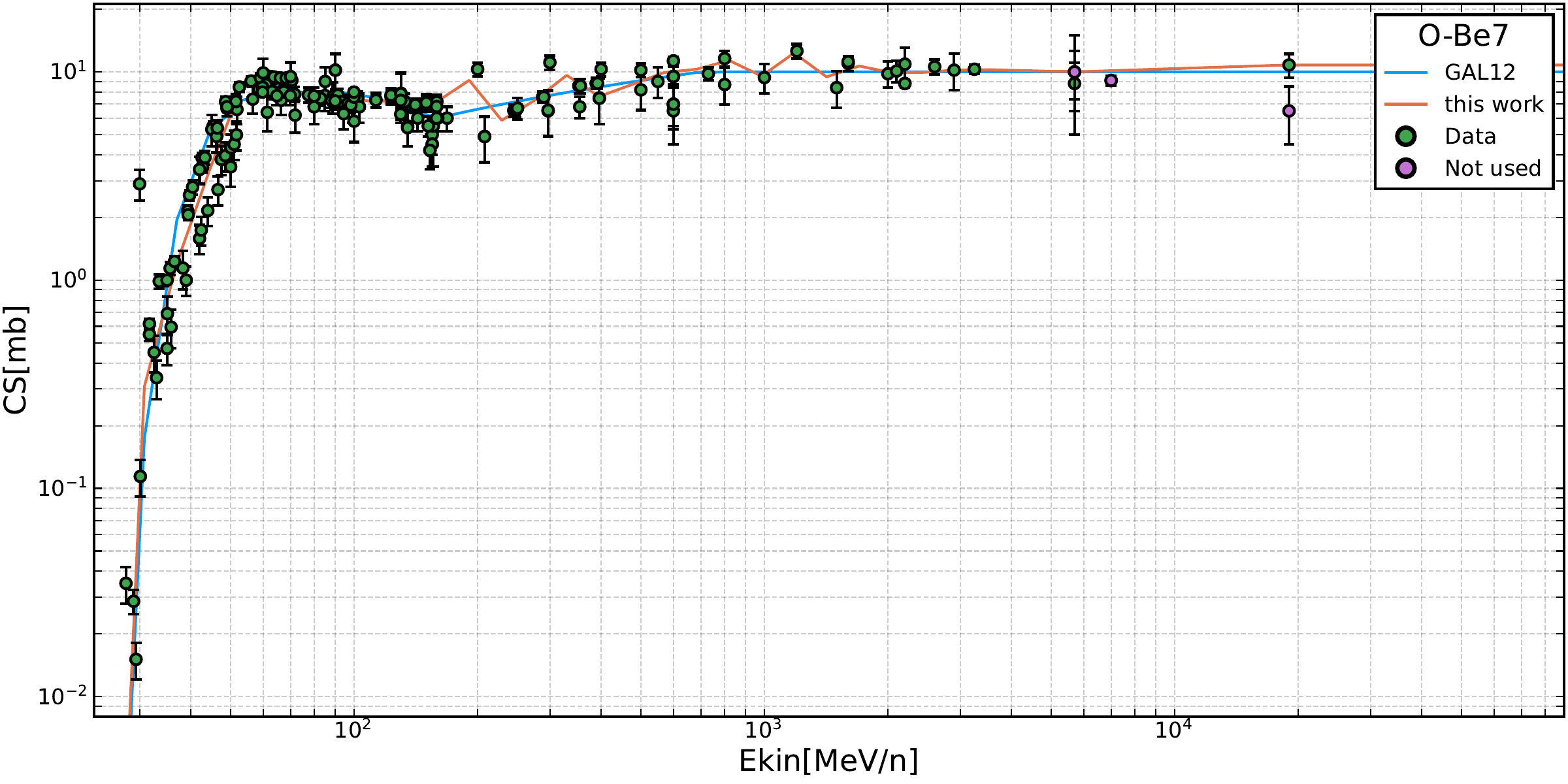}}
        \subfigure{  
           \includegraphics[width=0.45\textwidth]{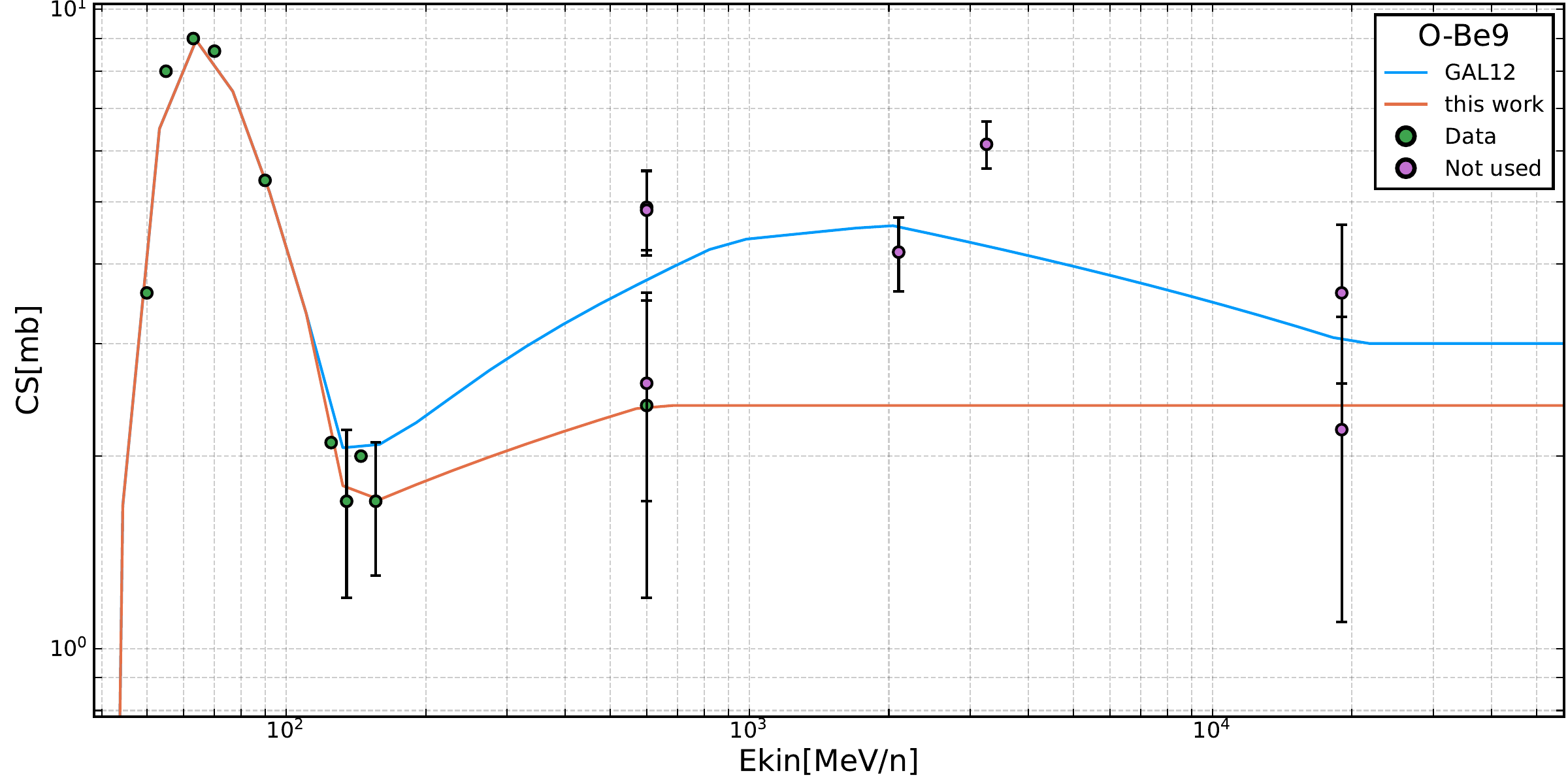}
      }\\
      \subfigure{
           \includegraphics[width=0.45\textwidth]{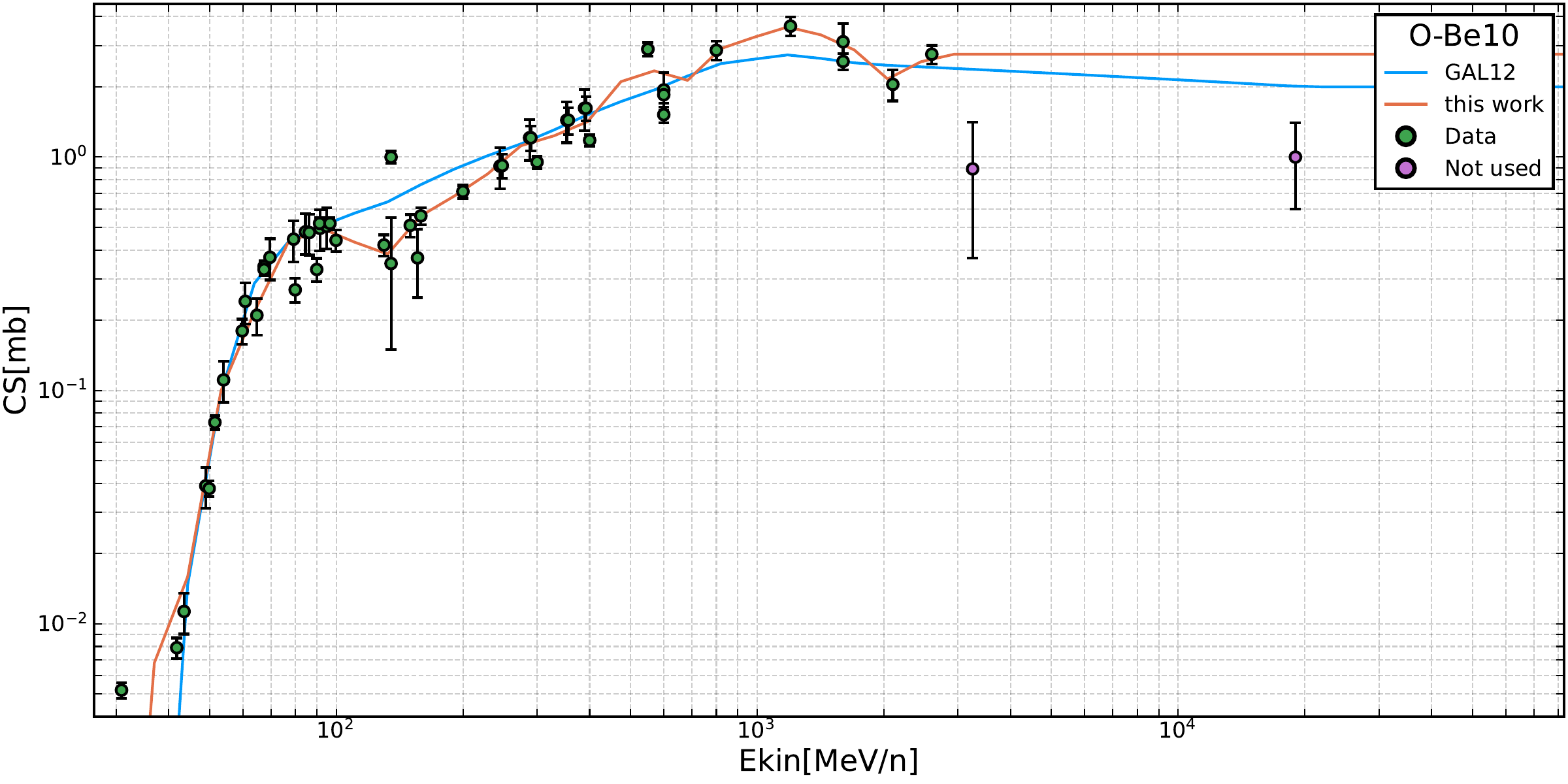}}
\end{figure*}

We have added thousands of data in \texttt{eval\_iso\_cs.dat} up to $\sim10$ GeVs to interpolate credible parametrization for important production channels of $\rm^2H$, $\rm^3He$, Li, Be, B, F, P, Sc, Ti and V.
As shown in Fig.~\ref{channel1} and \ref{channel2}, labeled as [Not used], some data points published by different groups could deviate from the general interpolation by a significant factor. We remove these measurements during the estimation of parametrization for a clearer determination. 
For example, the [KO99] and [KO02] \cite{Korejwo:2000pf, Korejwo:2002ts} data of several channels (such as $\rm^{12}\text{C} \longrightarrow ^{11}\text{B}$) were omitted as they significantly deviate from other measurements.
Sometimes we are not sure whether to discard an important measurement, which decides how the cross section becomes constant and extrapolates to the higher energy. For example, the newly measured data by [Ba05] \cite{Bazarov:2005qu} at 3550 MeV/$n$ indicates a smaller cross section compared with other measurements, and we cannot distinguish if the cross section reduction could happen at that energy. The parametrization of the channel $\rm^{16}\text{O} \longrightarrow ^{11}\text{B}$ predicts a reduction of nearly 30\%, and we choose to test its validity in the combined fitting together with CRs measurements in Sec.~\ref{sec:result}. The channel $\rm^{16}\text{O} \longrightarrow ^{10}\text{Be}$ predicts a reduction of over 70\%, hence we consider discarding it.
In Sec.~\ref{sec:be/b} we find that the channel $\rm^{16}\text{O} \longrightarrow ^{9}\text{Be}$ prefers a much lower cross section, owing to the expectation from the AMS-02 $\rm^{9}\text{Be}$ observation. Hence we omit almost all the points at above 1 GeV/$n$ and evaluate an energy-independent parametrization for this channel, which can naturally explain the $\rm^{9}\text{Be}$ measurement.

In these figures, we also illustrate the default parametrization [GAL12] defined in {\footnotesize GALPROP}. As designed in the evaluation routines,
the contributions of ghost nuclei $\rm^{10}\text{C}$ and $^{11}\text{C}$ are directly counted as the cumulative $\rm^{10}\text{B}$ and $\rm^{11}\text{B}$ if the projectile is $\rm{^{16}O}$ or $\rm{^{14}N}$. To compare [GAL12] and [GAL22] with other parametrizations and also the data, we subtract ghost nuclei based on Webber's prediction \cite{Webber:1998ex, Webber_2003} of the isotopic proportion to the cumulative cross section. This is the same as how G\'{e}nolini \cite{Genolini:2018ekk} treated ghost nuclei.
As a result, the default parametrization illustrated in some specific channels is labeled as [GAL12*] instead, representing the subtracted parametrization based on [GAL12]. The [GAL12] parametrization of the reaction $\rm^{16}\text{O} \longrightarrow ^{10}\text{B}$ illustrates the cumulative production $\rm^{10}\text{B}+^{10}\text{C}$ labeled as [GAL12(X10)], which fits the available measurements better and we didn't subtract it.

Illustrated in Fig.~\ref{channel1} and \ref{channel2}, several channels lack of high-energy ($>10~$GeV/$n$) measurements, such as $\rm^{12}\text{C} \longrightarrow ^{10}\text{C}$, $\rm^{16}\text{O} \longrightarrow ^{10}\text{B}$, $\rm^{16}\text{O} \longrightarrow ^{11}\text{B}$, $\rm^{16}\text{O} \longrightarrow ^{10}\text{C}$, $\rm^{16}\text{O} \longrightarrow ^{11}\text{C}$, $\rm^{16}\text{O} \longrightarrow ^{9}\text{Be}$ and $\rm^{16}\text{O} \longrightarrow ^{10}\text{Be}$. Most of them are less constrained with inadequate datasets, and we hope they could be improved by more cross-section observations in the future.

In addition, we attach ancillary files in the arXiv version that can be used to check the B and Be production results given in this work. These files (\texttt{eval\_iso\_cs.dat}, \texttt{isotope\_cs.dat} and \texttt{p\_cs\_fits.dat}) can be downloaded and added to the original files of the {\footnotesize GALPROP} code in the ``galtoolslib$\slash$ nuclei'' folder for testing purpose.

\end{document}